\documentclass[conference,compsoc]{IEEEtran}
\PassOptionsToPackage{hyphens}{url}

\def\isanonymous{0}
\def\issmallscreen{0}

\usepackage[l2tabu,orthodox]{nag}

\usepackage{ifxetex}

\usepackage{microtype}
\usepackage{booktabs,caption}
\usepackage{hyphenat}
\usepackage[flushleft]{threeparttable}
\usepackage{longtable}
\usepackage{array}
\usepackage{multirow}
\usepackage{lscape}
\usepackage{siunitx}
\usepackage{hyperref}

\usepackage{ifthen}
\newcommand{\anonymous}[2]{%
	\ifthenelse{\equal{\isanonymous}{1}}%
	{{#1}}%
	{{#2}}%
}

\newcommand{\smallscreen}[2]{%
	\ifthenelse{\equal{\issmallscreen}{1}}%
	{{#1}}%
	{{#2}}%
}

\ifthenelse{\equal{\issmallscreen}{1}}{
	\usepackage{geometry}
	\geometry{margin=5cm}
	\setlength{\marginparwidth}{2cm}
}{}

\usepackage{xcolor}
\definecolor{oxygenorange}{HTML}{FFDD00}
\usepackage[color=oxygenorange]{todonotes}

\ifthenelse{\equal{\isanonymous}{2}}
{
	\newcommand{\rikke}[2][]{}
	\newcommand{\taylor}[2][]{}
}{
	\newcommand{\rikke}[2][inline]{\todo[#1]{\textbf{rikke:} #2}\xspace}
	\newcommand{\taylor}[2][inline]{\todo[#1]{\textbf{taylor:} #2}\xspace}
}

\usepackage{amsmath,amsfonts}  
\usepackage{xspace}
\usepackage{cleveref}

\usepackage[lambda,landau,operators,probability,sets,logic,complexity,asymptotics]{cryptocode}

\usepackage{booktabs}  
\usepackage{comment}
\usepackage{enumitem}

\usepackage{navigator}
\usepackage{url}
\usepackage{graphicx}
\usepackage[export]{adjustbox}
\usepackage{float}
\usepackage{multirow}
\usepackage{longtable}
\usepackage{array}
\newcolumntype{L}{>{\raggedright\arraybackslash}p{3cm}}
\usepackage{pdfpages}
\usepackage{subcaption}   
\usepackage{tikz,pgfplots,pgfplotstable}
\usepgfplotslibrary{statistics}
\usetikzlibrary{calc}
\usetikzlibrary{arrows}
\usetikzlibrary{positioning}

\pgfplotsset{
	tick label style={font=\small},
	label style={font=\small},
	legend style={font=\small, cells={anchor=west}}
}

\definecolor{DarkPurple}{HTML}{332288}
\definecolor{DarkBlue}{HTML}{6699CC}
\definecolor{LightBlue}{HTML}{88CCEE}
\definecolor{DarkGreen}{HTML}{117733}
\definecolor{DarkRed}{HTML}{661100}
\definecolor{LightRed}{HTML}{CC6677}
\definecolor{LightPink}{HTML}{AA4466}
\definecolor{DarkPink}{HTML}{882255}
\definecolor{LightPurple}{HTML}{AA4499}

\definecolor{DarkBrown}{HTML}{604c38}
\definecolor{DarkTeal}{HTML}{23373b}
\definecolor{LightBrown}{HTML}{EB811B}
\definecolor{LightGreen}{HTML}{14B03D}

\usepackage{listings}
\lstdefinelanguage{Sage}[]{Python}{morekeywords={True,False,sage,cdef,cpdef,ctypedef,self},sensitive=true}
\begin{document}
	
	\title{Searching for a \emph{Farang}:\thanks{\emph{Farang} is a term used by Thai people to refer to white Western foreigners.}\\Collective Security among Women in Pattaya, Thailand}
	\anonymous{
		\author{}
	}{
		\author{
			{\rm Taylor Robinson}\\
			Royal Holloway, University of London\\
			taylor.robinson.2021@live.rhul.ac.uk
			\and
			{\rm Rikke Bjerg Jensen}\\
			Royal Holloway, University of London\\
			rikke.jensen@rhul.ac.uk			
		}
		
	}

	\clubpenalty=1
	\displaywidowpenalty=1
	\widowpenalty=1
	
	\maketitle

\begin{abstract}
We report on two months of ethnographic fieldwork in a women's centre in Pattaya, and interviews with 76 participants. Our findings, as they relate to digital security, show how (i) women in Pattaya, often working in the sex and massage industries, perceived relationships with \emph{farang} men as their best, and sometimes only, option to achieve security; (ii) the strategies used by the women to appeal to a \emph{farang} involved presenting themselves online, mirroring how they were being advertised by bar owners to attract customers; (iii) appealing to what they considered `Western ideals', the women sought out `Western technologies' and appropriated them for their benefit; (iv) the women navigated a series of online security risks, such as scams and abuse, which shaped their search for a \emph{farang}; (v) the women developed collective security through knowledge-sharing to protect themselves and each other in their search for a \emph{farang}. We situate our work in emerging digital security scholarship within marginalised contexts. 
\end{abstract} 

\section{Introduction}\label{sec:introduction} 

\noindent Pattaya, a tourist hub in Thailand, is often associated with its sex and hospitality industries~\cite{chiablaem:2020aNALYSIS,DMM:longjit2013,JAH:Phannarat:2016}, where many women from economically disadvantaged regions of Thailand, such as Isan, migrate in search of better opportunities and an improved income~\cite{cameron2005migration,JID:le2015}. Many of these women see pursuing a relationship with a Western foreigner, a \emph{farang}, as their best way to escape a life of poverty~\cite{APJMR:Pomsema2015,SIM:Inequality2019,PSP:Statham2021,Thailand:sunanta2014}. However, searching for and entering these relationships, while offering potential financial benefits, also exposes women to a spectrum of (digital) security risks -- ranging from scams through dating platforms and the marketing of women online by bar owners to physical security threats of abuse and trafficking. We show how the women in our study collectively developed protection strategies, often through online knowledge-sharing, against such threats. 

A growing body of security and privacy scholarship focuses on the digital security practices of different groups in what is often referred to as marginalised and/or at-risk contexts~\cite{warford2022sok}. This includes, for example, work on the performative practices of privacy by women in South Asia~\cite{sambasivan2019they}, the scapegoating of women (and other marginalised groups) in Lebanon~\cite{USENIX:MccJenTal23} and power dynamics and privacy in the use of smart technology within Jordanian households~\cite{USENIX:AlbFle23}, to mention a few. None to our knowledge have done so in the context of the Pattaya sex and massage industry, within which many of the women in our study found work. Our study therefore also draws upon emerging research that considers security and privacy in the context of sex work; much of which has been conducted in Western contexts, e.g.~\cite{barwulor2021disadvantaged,mcdonald2021s,CHI:SNBMAH24}, as also noted in~\cite{CHI:StrClaLai19} where the authors carried out fieldwork with sex workers at a brothel in Bangladesh. Although little research has been directed toward the sex industry in Thailand, or, more broadly, the Global South, accounts from a diversity of sources highlight the security risks facing sex workers in Thailand. This includes threats related to online sex work~\cite{Craig2024-ae,Hung2024-us} such as the trafficking of women forced into (online) sex work~\cite{Nguyen2025-no}, which grew considerably during and after COVID-19 as many Thai sex workers were without work~\cite{Duangdee2020-ke,Nguyen2025-no}.   

\subsubsection*{Contributions} We conducted ethnographic fieldwork over two months (January -- March 2024) in Pattaya, Thailand, following a one-month (May -- June 2023) scoping trip to Bangkok and Pattaya. This involved participant observation in women's centres in Pattaya, and individual and group interviews with 60 women who spent time in one of these centres, and 16 centre staff, volunteers and stakeholders. The conditions of Thai women in Pattaya -- many of whom had migrated from poorer regions in Thailand, had left abusive Thai partners, were over the age of 40 and now engaged in sex and massage work -- led many to seek the security of having a \emph{farang} partner.\\

\noindent  The aim of our research was to explore \textbf{how Thai women in Pattaya navigate a variety of (digital) security risks} in their daily lives. Within this aim, our work makes four distinct contributions to digital security.

\begin{itemize}
	\setlength\itemsep{0em} 
	\item The strategies adopted by the women in their search for a \emph{farang} often involved presenting themselves online in ways they considered appealing to Western men, while they sought out online platforms and technologies that they considered `Western' and appropriated them for their needs. This led to particular security risks, including threats, online scams, human trafficking.
	\item The women in our study developed knowledge-sharing practices grounded in collective (digital) security, which were often cultivated through relationships with other women at the centre and extended through online groups. Here, the women learned how to detect scammers and \emph{scan} for a \emph{good} \emph{farang}, while practising the use of image-alteration technology and generative AI tools to help them appeal to a \emph{farang}. 
	\item Computer and security education at the centre was aimed at the situated security needs of the women, rather than well-established computer security advice, and was taught by educators with lived experiences that resembled those of the women.
	\item Our contextually grounded findings have distinct implications for digital security research and education. For example, security for the women in our study was not rooted in the security of a piece of technology, but in how this piece of technology could benefit them in their particular situation. 
\end{itemize}

\noindent We develop these contributions in~\Cref{sec:discussion}. 

\section{Context and Related Work}\label{sec:related-work}

\noindent Pattaya is one of Thailand's top tourist destinations for foreigners~\cite{DMM:longjit2013} with an estimated 27 million visitors in 2024~\cite{PRD:Pattaya2024}. The city experienced rapid and unplanned growth in the 1960s and 1970s due to the presence of the US military in Vietnam, which also fuelled the prevalent sex industry Pattaya is known for today~\cite{DMM:longjit2013,JAH:Phannarat:2016}. Migration into tourist hotspots such as Pattaya is perceived to provide rural and poor Thai families increased income~\cite{JID:le2015}, with women leaving to work and send money back to their families. This was the case for most of the women in our study.

\subsection{The (In)Securities of Sex Work}\label{sec:setting}

\noindent In Pattaya, many women take up sex and massage work as these are typically the highest paid jobs for those with low education levels. Thus, while our research did not specifically focus on sex work, most women in our study also worked or had worked in the sex industry in Pattaya. This often left them isolated and stigmatised in a city that was largely alien to them. They competed with other women for clients (\emph{farang} men) and had to satisfy bar owners and \emph{mamasans}.\footnote{The \emph{mamasan} often came from the industry herself and, e.g.~managed the working hours of the women at a bar and sex workers' behaviours~\cite{SD:plambech2023}.} This included providing regular blood and urine samples to bar owners, where a positive STD test result would lead to them losing their work. Stories of women being killed or trafficked by a client were frequent, while most women working in the sex industry had experienced abuse and threats. 

\subsubsection*{Online Sex Work and Security} Sex work is considered illegal in Thailand and, as a result, sex workers are largely invisible within the formal Thai economy and are offered little to no societal protection. This was particularly evident during the COVID-19 pandemic where Pattaya's tourism almost came to a halt~\cite{hasayotin2024empowerment}; this put many women, who worked in the sex industry in Pattaya and relied on clients from abroad, out of work. Sex workers, unlike other `formal' sectors, received no economic support from the Thai Government and had to rely on family support or personal savings~\cite{bishop2024bar}. Moving client engagements online became a way for many women to increase their earnings, while also providing them with a form of protection from physical abuse. As others have found in Western contexts (e.g.~\cite{barwulor2021disadvantaged,mcdonald2021s}), online sex work cuts out the `middle person', enabling women to negotiate with clients and remove costs associated with in-person meetings. Online sex work has also been shown to increase privacy through greater anonymity~\cite{barwulor2021disadvantaged,mcdonald2021s}, reducing the impact of social stigma and, in Thai and other Asian contexts, the risk of \emph{losing face}~\cite{veena2007revisiting}.\footnote{\emph{Losing face} refers to damaging one's reputation or credibility, often leading to shame or social exclusion, while \emph{saving face} includes acting in a way that preserves dignity and honour.}

A growing body of security and HCI research has examined the impact of online platforms on the sex industry across diverse national contexts. For example, the authors of~\cite{CSCW:HamBarRed22} focused on the impact of COVID-19 on sex workers in the Global North, while others have explored how sex workers find clients online~\cite{barwulor2021disadvantaged,castle2008ordering,veena2007revisiting} and how they use digital tools to engage in video-based sex work and advertise themselves online~\cite{sanders2016our,scoular2019beyond}. Much of this existing work highlights how online platforms give sex workers more control over who their clients are and how they advertise their services, resulting in some having higher earnings than those who engage in more `traditional' sex work~\cite{barwulor2021disadvantaged,castle2008ordering,sanders2016our,scoular2019beyond}. Being able to negotiate with clients online has also been shown to improve the safety of sex workers, helping them reduce physical and psychological risks of in-person sex work, including violence, coercion and unsafe sex~\cite{barwulor2021disadvantaged}. 

Recent security-driven work has explored the specific digital security needs of sex workers. For example, some studies have focused on the extent to which sex workers use digital networks to protect themselves, such as through sharing client information and verifying client identities face-to-face~\cite{barwulor2021disadvantaged,warford2022sok}. The authors of~\cite{CHI:StrClaLai19} and~\cite{CHI:StrLaiCom17} explored how organisations that focus on sex workers' rights use digital technology to facilitate knowledge and information sharing to provide greater protection for sex workers. Others have highlighted how sex workers manage their digital identities through separate or anonymous accounts to safeguard personal information~\cite{barwulor2021disadvantaged,mcdonald2021s,sambasivan2019they}. For example, the authors of~\cite{mcdonald2021s} studied the digital security needs of sex workers in a European context. While they found that sex workers were already skilled at managing their online security, technologies often failed to support their specific needs. To manage risks associated with sex work, participants would develop strategies such as having a device for sex work and one for personal use, avoiding identifiable photos and `vetting' clients using online networks. 

Much of this work, however, is focused on Global North contexts, e.g.~Europe~\cite{mcdonald2021s}, Canada~\cite{CHI:StrClaLai19}, UK~\cite{CHI:StrLaiCom17}, Germany and Switzerland~\cite{barwulor2021disadvantaged}, where sex work is partly legalised and regulated, less stigmatised than in some Global South contexts -- and where many sex workers might have greater access to and are more familiar with digital technology and its security. As noted by the authors of~\cite{CHI:SNBMAH24}, much existing HCI literature (and security-driven scholarship, as we show here) on the intersection of sex work and digital technology does not consider the Global South, although sex workers in this region represent a large portion of the global sex industry.\footnote{In~\cite{USENIX:HasDaiAki24}, the authors make a broader claim about usable privacy and security research being particularly WEIRD (Western, Educated, Industrialised, Rich, Democracy); following the work in~\cite{CHI:LSBCOR21}.} Through an ethnographic study in a brothel in Bangladesh, they show how sex workers protected themselves from security threats while trying to overcome their lack of technological know-how through shared and creative practices. Our study contributes to this work. 

\subsection{Marginalisation and Security}\label{sec:rw-marginalisation}

\noindent We further contextualise our contributions by drawing on the growing body of security-driven work that has focused on groups experiencing different forms of marginalisation, recognising that their security needs are not well served by the technologies they rely upon. The participants in our study can be considered \emph{marginalised} in multiple ways, grounded in their lived experiences of deep-rooted economic hardship, abusive intimate and work relations, as well as social stigma linked to their (sex and massage) work in Pattaya. Their marginalisation further shapes their use of and reliance on digital technology, while intensifying their individual and shared digital security needs. 

The intertwining of technological and social dimensions of security is exemplified in existing work with marginalised groups. For example, the authors of~\cite{sambasivan2019they} showed how South Asian women developed performative practices to protect their privacy from intimate and familial relations. In~\cite{USENIX:MccJenTal23} the authors highlighted how marginalised groups in Lebanon experienced being scapegoated for wider societal-wide failures, focusing on groups identifying as LGBTQIA+ and refugees as well as women who spoke out about their marginalisation. This work was followed up in~\cite{CSCW:MccJenTal24}, where the authors demonstrated how collective security practices helped participants `patch' security for themselves and others in the face of failing infrastructures. In~\cite{USENIX:ABJM21} the authors showed how the idea of collective security was central to Anti-ELAB protesters in Hong Kong. Tying together notions of financial insecurity and technological security, the authors of~\cite{CHI:SMOTWSOSC19} explored how pressures relating to homelessness and limited financial resources impacted people's security and privacy practices. Thus, the recognition that marginalisation, rooted in distinct social and societal contexts, has implications for digital security is well-established in current scholarship. 

The literature further highlights the diverse contexts within which marginalisation is experienced and its implications for security. In particular, and as we observed through our present study, marginalised communities often have to rely on the security and technological knowledge held by support networks and social relations, creating a series of dependencies. For example, research on the protective mechanisms developed by migrant domestic workers highlights how these were grounded in wide-reaching support networks~\cite{USENIX:SCBAPB22}. Much security-focused research on refugees makes similar findings, where socially-rooted barriers such as language, financial resources and cultural knowledge have been found to establish a dependency on others~\cite{CHI:ColJenTal18,CHI:JenColTal20,SP:SLIRK18}. Recent work on LGBTQIA+ people also showed how they looked for support from trusted queer groups to navigate questions of identity, personal safety and security~\cite{USENIX:GHRR22,CHI:LHKZH20}. 

Other forms of marginalisation considered in the security literature also intersect with the context under study in our work. These relate to distinct security risks materialising through intimate and abusive relations. For example, in~\cite{CHI:MOTSWSMCC17}, the authors demonstrated how survivors of intimate partner abuse developed distinct privacy practices that pertained to the different stages of their experiences of abuse. This work speaks to a broader body of scholarship that has focused on the security and privacy needs of those living with intimate partner violence, e.g.~\cite{CSCW:FHTGCRD19,CSCW:FPMLRD17,CHI:TFERD21}. 

These works shine a light on the diversity of ways in which marginalisation intersects with security and technology, while it shapes how digital security is practised in people's daily lives. We now turn to everyday security in~\Cref{sec:rw-sen}.

\subsection{Security in the Mundane}\label{sec:rw-sen}

\noindent Everyday security considers security as a fundamental part of daily life~\cite{crawford2016mapping,nyman2021everyday}, comprising ``micro practices'' of security interactions between people, groups and institutions~\cite{crawford2016mapping,huysmans2006politics}. In line with~\cite{nyman2021everyday}, we define everyday security as consisting of three overlapping dimensions: (1) spatial (mundane spaces outside formal politics), (2) temporal (routines shaping security) and (3) affective (lived experiences of insecurity). Further, we draw on broader sociological ideas related to the creation of routinised daily activities to establish a sense of predictability -- and, by extension, (ontological) security\footnote{In the sociological literature, ontological security encompasses both the freedom to live without fear and protection from harm; ontological security then is the feeling that one's place in the world is stable and reliable and, thus, provides a sense of predictability~\cite{SII:mcsweeney1999,VPS:roe2008}.} -- in daily life~\cite{dupuis1998home,giddens2016modernity,phipps2017routines}. 

Everyday security shifts focus away from elite-driven security perspectives to the ways in which people create, maintain and disrupt security on a daily basis for themselves and others~\cite{eschle2018nuclear}. This focuses attention on the social, relational and intimate, rather than wider societal dimensions of security. For example, prior research highlights how marginalised groups build security through routine interactions and informal networks~\cite{gunning2022you,jones2019security,USENIX:MccJenTal23}, rather than through institutionalised forms of protection as these are often not available to them. Indeed, state provisions of security often create insecurities for marginalised or local communities, where the state security is placed at a higher value than that of local people, as noted in~\cite{eschle2018nuclear,higate2010space}. 

We are not the first to consider everyday security in research within marginalised contexts, while the need for consistency and routine to establish security in one's everyday life is increasingly recognised in security scholarship~\cite{joque2020deconstructing}. For example, the authors of~\cite{CSCW:MccJenTal24} showed how marginalised groups in post-conflict Lebanon developed distinct practices and routines to overcome infrastructural failures to maintain a sense of `normalcy' and control. In~\cite{CHI:colesJensen2019} and~\cite{CHI:ColJenTal18}, the authors considered how a broader (sociological) conception of security shone a light on the everyday practices and routines through which refugees and migrants to Sweden (re)established a sense of security in their new land.

We develop the idea of everyday security in~\Cref{sec:discussion} as it pertains to our findings in~\Cref{sec:findings}.

\section{Research Design}\label{sec:research-design}

\noindent The research was designed following a one-month scoping trip to Bangkok and Pattaya, to ensure that it was sensitive to the specific context under study. Here, the main author (hereafter: \emph{the fieldworker}) spent time engaging with people throughout Pattaya, including public events, walking streets\footnote{Streets where bars offering sexual services were most prominent} and community visits. Yet, most of the time was spent engaging with different women's centres supporting low-income women and children, including rehabilitation centres for women working in the sex industry, Christian missionary centres and centres providing education for women across Pattaya. These centres became spaces of observation and interaction with women visiting the centres for different reasons. They also provided the opportunity to engage other stakeholders, including staff, volunteers, donors and partners.

\subsection{Methods}\label{sec:methods}

\noindent The fieldworker conducted a two-month ethnography in Pattaya from January to March 2024. This involved participant observation in women's centres, and individual and group interviews with women in one of these centres. We refer to this as \emph{the centre} throughout. 

\subsubsection{The Centre} 

The stated mission of the specific centre is to help women in Pattaya, particularly through outreach initiatives aimed at encouraging women to leave sex work, while advocating for ending violence against women. However, the main function of the centre is to provide language classes (German, French and English) to local women, taught primarily by Western volunteers. Classes take place Monday to Friday, with morning and afternoon sessions, and lunch served in between. The centre also provides salon and massage classes, and runs shorter programmes on technology, such as a ten-day computer class and a two-day mobile phone class. A 30-minute weekly `town hall' includes dancing, announcements and broader topics such as climate change or mental health. During these sessions, it is very common to have visitors/donors as `honourable guests'. The centre also provides individual sessions on various social and legal issues facing the women, such as access to a counsellor, HIV testing, self-defence training and legal advice. 

\subsubsection{Participant Observation} 

Participant observation at the centre took several forms. First, the fieldworker attended women's classes each morning and afternoon, observing the learning approach and the women's emotional responses to what they were being taught -- as well as engaging in basic conversation with some of the women. Second, the fieldworker observed interactions between classes, such as during lunch, focusing on the interactions between the women and the centre volunteers and staff. During lunch, it was also customary for volunteers and staff to eat separately from the main dining area, which often resulted in conversations about what had occurred in other classes and conversations with the centre director. Third, observations took place during the Wednesday town hall sessions. This allowed the fieldworker to observe interactions between the women, volunteers and staff. Donors often attended the Wednesday meetings, enabling observations of how donors to the centre engaged with the women. Observations were recorded as full field notes and brought into the analysis. 

\subsubsection{Individual and Group Interviews} 

The research involved interviews with 76 participants: 60 Thai women living and working in Pattaya, and 16 stakeholders such as centre staff, volunteers and missionaries (see Appendix~\ref{app:participant-tables}). Participants were recruited through daily interactions with the women in the centre and no selection criteria, beyond participants having to identify as a woman, be over the age of 18 and attend one of the women's centres in Pattaya, were applied. Everyone wanting to take part was given a choice of either joining a group or individual interview to ensure that they could voice their thoughts in a way that best suited them. While the first few participants were recruited by the fieldworker approaching women attending daily classes in the centre, eventually women started to come forward after they had heard other women talk about having participated. 

Interviews with the women were conducted in a private room at the centre to preserve confidentiality. They were semi-structured and followed a topic guide (Appendix~\ref{app:interview-guide}), which was developed in consultation with the centre to ensure that the topics were as sensitive to the context as possible. The interviews were conducted in English or Thai, depending on the participants' preference and language ability. A translator was present for all interviews with the Thai women to ensure complete understanding of the consent process and interview questions. The translator signed a confidentiality agreement before starting the research process to ensure confidentiality and anonymity throughout the interview process. For interviews with the various stakeholders, who provided contextual information about the experiences of local Thai women, the majority spoke English and interviews were conducted in spaces where participants felt most comfortable (usually within a private room in one of the women's centres in Pattaya). Interviews lasted between 21 and 109 minutes (Appendix~\ref{app:participant-tables}) and were audio-recorded with the informed consent of each participant (\Cref{sec:ethics}), and later transcribed. 

\subsection{Ethical Considerations}\label{sec:ethics}

\noindent The research received full ethics approval from our institution's Research Ethics Committee (REC) and procedures for obtaining informed consent were meticulously followed. All participation was voluntary and participants were fully informed of the expectations of the research as well as their involvement in it. In line with our institutional ethical protocols and local guidance, participants were not financially compensated for taking part in the research. This was to avoid anyone feeling compelled to take part due to a financial incentive. Participants were provided with a detailed participant information sheet which provided details on the purpose of the research, data collection and storage, and information regarding general questions asked by participants (e.g. what questions will I be asked, etc.). They were also provided with a consent form in advance of their involvement, and these were provided in both Thai and English. At the start of any interview, the translator ensured that the information was fully understood. 

At the start of the fieldwork, a large group meeting was called at the centre, which was open to everyone. Here, the research and the fieldworker's presence were explained and women using the centre were able to ask questions and engage with the fieldworker in an informal setting. The research was also presented at the weekly group meetings and through the staff volunteers when women came to the centre. The fieldworker also visited each classroom daily, explaining the project and informing women of the opportunity to participate. No personally identifiable information was captured about any individual, while demographic data such as age is not linked to any participant. Following each interview, the recording was transcribed and anonymised. Interviews, transcripts, field notes and consent forms were digitised and uploaded to an institutionally approved server before leaving the field to avoid travelling across international borders with sensitive data. 

Many of the participants had previously experienced traumatic events, such as domestic abuse or sexual exploitation. Care was taken to avoid traumatic or triggering questions and discussions. Participants were also assured that if they felt uncomfortable at any point, they could leave or ask that the questions be readjusted to mitigate discomfort. Participants were reminded to seek counselling (free through the centre) for care and support, especially if they were upset or triggered by any part of the research process. The fieldworker ensured that potential risks were communicated with centre staff and arranged for counsellors to be available to discuss any experiences of participation right after the interview, as well as at any other time in the future. No participant asked to use these services after the interview. 

Given the sensitive nature of this work, we are unable to provide access to our full dataset. This is in line with established data protection practices in qualitative research, especially in contexts where sharing such data could risk participant identification and cause harm to them and/or their environments. Instead, we use pseudonymised and carefully curated quotes to support our findings and to ensure that the participants' voices remain present in our work. We also make our topic guide, participant tables, participant information sheet and reflexive thematic analysis table available as appendices.

\subsubsection{Researcher Wellbeing}\label{sec:researcher-wellbeing}

The research team comprises significant fieldwork experience in settings that can be considered higher-risk and marginalised in different ways. We drew on this experience throughout the research, while also working closely with the fieldworker's institutional H\&S office. We did so to establish appropriate safety protocols for the specific context, which included physical safety and emotional wellbeing practices and mitigations. This also covered considerations related to lone-working in the environment of Pattaya, for example. We implemented check-in procedures, where the fieldworker had contact with the rest of the research team at regular intervals; this was done through both messaging and calls. We also had a course of action in case the fieldworker missed a check-in (this was, however, never the case). We established these protocols before the start of any fieldwork and iteratively re-considered them throughout the fieldwork. The fieldworker also engaged in practices such as note-taking and journaling which included self-reflections on their own emotional state, while they established daily routines that also included free time and decompression. We also had access to a wellbeing team through the fieldworker's institution, although we never needed to involve them.

\subsection{Data Analysis}\label{sec:data-analysis}

\noindent Our data analysis was designed to ensure that our interpretations remained grounded in the settings we studied as well as participants' voices and experiences. Data was analysed inductively, following principles outlined by Hammersley and Atkinson~\cite{EPP:hammersley2019}, who emphasise the need to stay with the naturalistic, interpretive, iterative and reflexive nature of fieldwork. Staying with the ethnographic also means avoiding claims about \emph{saturation} or \emph{inter-rater reliability}, which appeal to a more neo-positivist, rather than interpretative  (see e.g~\cite{QESEH:BraCla21} and ethnographic writings such as~\cite{WEF:emerson2011}), approach. The themes we present in this work (\Cref{sec:findings}) thus emerged directly from the data through several rounds of analytical coding. 

\subsubsection{Coding and Themes}\label{sec:coding}

We started our analysis with the fieldworker digitising field notes and transcribing all interviews, before reviewing this data to identify recurring or significant themes~\cite{WEF:emerson2011}. Our first coding round involved open coding in a colour-coded Excel spreadsheet, where the fieldworker grouped quotes and notes into initial codes such as, for example, the code \emph{`authentication of the \emph{farang}'} (labelled `Codes' in Appendix \ref{app:reflexive-analysis}). As new patterns emerged, codes were updated and refined. The initial themes were discussed within the research team through extensive analytical sessions. In the second coding round, the fieldworker used selective open coding, focusing on key themes from the first round and collaborative analysis, while remaining open to new insights~\cite{WEF:emerson2011}. Since we compartmentalised within the research team to ensure that only the fieldworker had access to the raw data, the fieldworker constructed preliminary summaries of key themes and reflections to guide collaborative discussions within the research team (see `Descriptive/interpretive summary (round 1)'). This process, along with collaborative discussion among the authors, helped refine codes and interpretations and shape the core themes for more focused analysis. This helped us refine the main themes, construct sub-themes and map connections between them. We updated our Excel coding tables to reflect this deeper analysis, combining earlier codes into broader themes like \emph{`Collective Security'} and exploring related subcategories (e.g.~protection from scammers)~\cite{QRP:braun2021,RH:braun2016,ALQ:muir2023} (see `Initial reflexivity constructed category'). Similar to the open coding stage, this stage also included updated and refined summaries to gain deeper insights into the themes (see `Interpretive summary (round 2, group analysis'). During the final stage of the analysis, we collaboratively drafted our findings in line with the themes that emerged from the analytical process (see `Final reflexively constructed theme, collaborative write-up'). Each draft went through multiple rounds of reflection, which led us to often revisit and restructure our codes. Throughout this process, we stayed with the complexity and diversity of participants' lived experiences as we observed in our data.

\subsubsection{Researcher Positionalities}\label{sec:researcher-positionalities}

Our individual and shared positionalities shaped our analysis of the data, as is common practice when adopting an interpretivist positioning. For example, field notes were both records and interpretations of meaning~\cite{WEF:emerson2011}, and we engaged in continuous reflection to refine emerging themes~\cite{ALQ:muir2023}. Recognising that this positioning shaped coding, we remained reflexive in developing themes~\cite{QR:trainor2021}. All fieldwork was conducted by the main author, a North American woman who also undertook the one-month scoping trip to Thailand. Another author, a European woman, uses ethnography to study information security practices in high-risk contexts. Together, we analysed the data through our individual and collective positioning~\cite{braun2022starting}. Both researchers have experience of researching security within underserved groups, but do not share the lived realities of the participants. Prior studies on Thai women often assume they are vulnerable, poorly educated and face social stigmas; assumptions to which we do not subscribe. The women in our study were, in many ways, resourceful. 

\subsection{Limitations} 

\noindent First, we involved a translator to help translate most of the interviews because the majority of the participants were not fluent in English. Although the fieldworker tried to account for language and communication barriers, nuances in language, meaning and cultural cues may have been lost or misinterpreted. Second, all interviews were conducted within a specific setting, the centre, grounding the data in a small community of women attending the centre and may not be fully applicable to the experiences of women throughout Pattaya or, more broadly, Thailand. Yet, given the inherently qualitative nature of this work, we consider this a strength of our approach. Further, ethical considerations, such as confidentiality, may have influenced the information that participants shared during interviews. Some participants may have censored their responses because of the sensitivity of their personal experiences. Finally, as with all (qualitative) research, our individual and collective positionalities shape our interpretations of findings. Yet, as is inherent in ethnography, we conducted our analysis reflexively.

\section{Findings}\label{sec:findings}

\noindent Many of the participants discussed previous experiences of domestic abuse from Thai ex-partners, single-motherhood and working in jobs requiring hard physical labour, and saw seeking work in Pattaya as a way to change their situation. However, many found Pattaya a difficult and dangerous place to work, leading them to look for a \emph{farang} partnership. \emph{Farang} relationships have become more common in recent decades not least due to accessible long-distance travel, mobile phones, messaging applications, internet dating and international money transfers. Such developments enable \emph{farangs} to frequently visit Thailand while allowing Thai women to connect with interested \emph{farangs}~\cite{ijir:McKenzie2021,PSP:Statham2021}. While the men are typically from Western Europe, Scandinavia, Australia or the US~\cite{PSP:Statham2021,Thailand:sunanta2014}, the women are usually from Isan~\cite{ijir:McKenzie2021,Thailand:sunanta2014}. Marrying a Thai woman provides the men with domestic and caring support~\cite{APJMR:Pomsema2015,PSP:Statham2021}, while marrying a \emph{farang}, known as becoming \emph{mia farang} (white Westerner's wife), is seen as a path out of poverty and improved social mobility for many Thai women~\cite{APJMR:Pomsema2015,PSP:Statham2021,Thailand:sunanta2014}. 

We present our findings in line with the themes constructed through the analysis detailed in~\Cref{sec:data-analysis}, with a focus on the steps taken by the women in our study to secure a \emph{farang} partner. To understand why the women in our study saw relationships with \emph{farang} men as a path toward security, we first discuss the everyday insecurities they faced in Pattaya in~\Cref{sec:Insecurities_Pattaya}. In the subsequent section, \Cref{sec:attracting-farang-online}, we present findings related to the digital activities and strategies of the women in their search for a \emph{farang}. In~\Cref{sec:collective-security} we show how the women in our study developed collective practices and shared information through (online) networks to protect themselves from (digital) security risks. 

\subsection{The Insecurities of Working in Pattaya}\label{sec:Insecurities_Pattaya}

\noindent This section introduces the shame and associated risks that the women often experienced from working in Pattaya. These risks were interwoven with their experiences of searching for \emph{farangs} online and through the online marketisation by both bar owners and themselves.

\subsubsection{\emph{Losing Face} in Pattaya} \label{sec:Losing_Face}

Many women came to Pattaya out of financial necessity, as jobs in the sex and massage industries offered higher pay than agriculture, factory or domestic work. Still, many women described the shame of working in Pattaya, a place widely viewed as a ``sin city''. For example, P1, a 53-year-old from Udon Thani, described the hurt of hearing her son say: \emph{``Mum, why did you go to Pattaya? That city is sin city!''} Hence, many women would withhold information about where they worked from their families to avoid \emph{losing face}. P59, a 46-year-old from Isan, explained: \emph{``It is dishonourable for a woman to live in Pattaya. Your reputation is ruined, and people assume you work in a bar. I only told my sister once my mum passed two years ago.''} Others shared that even when they had not \emph{lost face}, the association with Pattaya was enough to strain family relations because of the assumption that they were sex workers. This led to some women being cut off from their family and children, even if they had initially moved to Pattaya to financially support them. This led many to feel exploited or reduced to being a source of income. P29, a 46-year-old woman from Isan, stated: \emph{``Nobody supports me. I support myself. The only thing I'm asked [by my family] is `when will I send money?' They pressure me all the time.''} 

In addition to family pressure and shame, participants voiced how they faced a series of workplace threats in the sex and massage industries: exposure to HIV and other STDs, physical violence and sometimes death from going with the wrong customer. \emph{``You see the girls beaten up and lying on the street from going with the wrong man''}, P57, a 57-year-old from Udon Thani, noted. Many faced sudden job loss without any protection. Both massage and sex work environments were often described as toxic and highly competitive, with women accusing each other of stealing clients or engaging in workplace bullying, which often caused declining mental health. P33, a 40-year-old woman from Udon Thani, expressed: \emph{``They started bullying me in the group and spoke behind my back [\ldots] If there were 40 people in the parlour, 30 were bitching about me and talking behind my back. It was too much for me [\ldots] [I] couldn't take it any more.''} This was exacerbated with many women living in the bars or parlours where they worked, leaving them with no privacy or personal protection. Simply being in Pattaya, regardless of their work, exposed women to harassment from tourists who assumed they were sex workers, leaving many feeling unsafe even outside of work.  

For many women, the reality they encountered in Pattaya, including \emph{losing face} within their families and significant risks in their workplaces, was not what they had anticipated from migrating to Pattaya. Many viewed their move to Pattaya as a personal sacrifice, to provide increased financial security and improved opportunities for their family and children. E2, a North American expat working with various charities in Pattaya, explained: \emph{``Everybody in the villages knows that, or thinks they know, that it is lucrative to have women come to Pattaya and be in the bar.''} 

Ultimately, the women had a vision that moving to Pattaya would bring them a better life, but instead they found themselves isolated, ashamed and in an alien city. Having \emph{lost face}, being excluded and isolated from their families, left them without a familial network that they could turn to for support. As P34, a 41-year-old from Bueng Kan, stated: \emph{``A friend of mine invited me to come to Pattaya. I was only 24 and I had a dream in my head [\ldots] Someone told me to work at a GoGo bar [\ldots] But it was discouraging and I felt so sad. I was depressed.''} The experience of P34 resembles that of many women in our study arriving in Pattaya. It shines a light on how the women, on the one hand, encountered a series of threats they had never previously experienced while being forced to live in their place of work (bars and massage parlours) and, on the other hand, were isolated from their familial support networks due to the shame that came with being a woman in Pattaya.

Many women, thus, developed their own strategies for reducing the risks they faced while working with customers. Some participants working as sex workers attempted to reduce the threats they faced by staying in areas they knew had CCTV cameras, while others would also strike deals with specific hotels throughout Pattaya. Here, the women would agree to bring their clients to a particular hotel in exchange for a room for a few hours. Hotel staff would check on the woman every few hours to ensure her safety. These practices were often a way for the women to establish some form of safety and control. 

\subsubsection{Shame and the Marketing of Women Online} \label{sec:Marketing_Women}

Many women were marketed online by bar owners or \emph{mamasans}, usually without their consent and with no control over how they were being used. Photographs were taken weekly and posted to social media platforms like Facebook and TikTok. These images often depicted women in sexualised poses or costumes designed to attract clients. V4, a volunteer at the centre, described: \emph{``Most don't know how the photos are used, but I am sure they see on Facebook how they are depicted [\dots] It's the `land of smiles' but you don't see a lot of smiles.''} Through this form of advertising, bar owners or \emph{mamasans} would remain in control of the photos and, thus, how the women were presented -- both in the street and online -- as they decided their outfits and marketing approach. Observations during the fieldwork further underscored this with thousands of women lining the walking streets, dressed in various costumes representing many different sexual fantasies and desires. While engaging in the sex industry, many women often felt pressured to follow the directives of the \emph{mamasan}, often a matronly figure who managed a team of women, such as sex workers, working at a bar. 

The photos used to market the women and their services were observed to lead to severe consequences for some women. For example, many participants voiced concerns over being located in Pattaya by abusive ex-partners and controlling relationships and talked about how these photos placed them at a particular bar during a particular period. Participants voiced concerns that, if discovered, such images could also result in them \emph{losing face} within their local community and/or with their family. Others noted how photos of them presenting as sex workers could limit their future prospect of leaving the sex industry, with concerns being raised about their ability to secure jobs or foreign visas. This was also seen to make them potentially less attractive to \emph{farangs}. Importantly, many women lacked the know-how to find, edit or remove these images. As S1, a technology teacher at the centre, recalled: \emph{``One woman saw how public the images were [\ldots] She became worried and wanted to delete it, but I told them they couldn't.''} These images posed risks to women’s social and economic futures, and they frequently had no control over how they were used or the resulting consequences from such images being distributed. 

\subsection{Searching for a \emph{Farang} Partner Online}\label{sec:attracting-farang-online} 

\noindent While \Cref{sec:Insecurities_Pattaya} focused on the insecurities that women faced while living and working in Pattaya, this section reports on how and why the women in our study searched for a \emph{farang} partner. \Cref{sec:Benefits_Farang} highlights the perceived benefits of a \emph{farang}, while \Cref{sec:presenting-online} shows how participants leveraged digital technology to attract a \emph{farang}. Our study shows that it was through these online platforms that a Thai woman would present herself to a \emph{farang} and where conversations between a Thai woman and a \emph{farang} man would be initiated. However, as we show in~\Cref{sec:protection-physical-harm}, our work also points to the insecurities that the women felt in engaging with \emph{farang} men online.

\subsubsection{Perceived Benefits of a \emph{Farang}}\label{sec:Benefits_Farang}

\noindent Compared to the instability of local jobs in Pattaya, relationships with \emph{farang} men were seen to provide increased security. As P24, a 50-year-old from Udon Thani, highlighted: \emph{``I was working many different jobs, I was tired. Then I saw our neighbour, she has a foreign husband and it looked like her life was getting better and better.''} Others explained that when a Thai woman began dating a \emph{farang}, it was expected that in return for taking care of domestic duties and partnership, the \emph{farang} would pay the Thai woman an agreed-upon weekly or monthly fee. This fee would help support herself and her family, with fees ranging from several hundred to several thousand baht per week. While participants did not equate these relationships with sex work, they acknowledged their transactional nature. P28, a 46-year-old from Isan, reflected: \emph{``If we had a choice, I'd rather find money and support ourself. I wish I could stand on our own two feet, without a man. But support from a foreigner can change your life.''}

Some women viewed \emph{farang} relationships as a chance to gain skills or migrate abroad, reducing their reliance on sex work. P17, a 29-year-old from Isan, who had a Norwegian boyfriend, stated: \emph{``My boyfriend helps me financially with everything and whatever I want to study [\ldots] He supports me because if we go back to Norway, I won't have to go and start work from zero and start cleaning toilets.''} Some women saw \emph{farang} partnerships as an escape from past abuse and social limitations. P12, a 40-year-old woman from Isan, recalled: \emph{``The foreign men don't care if you have kids or what passport you have, they generally support you.''} Women who had experienced abusive Thai relationships and financial hardship hoped \emph{farang} men would provide both stability and emotional support. These beliefs were often reinforced by idealised depictions of Western relations through various forms of Western (social) media. They were also reinforced by conversations women had with Western volunteers at the centre. Many women believed that \emph{farang} men wanted a woman who would embrace traditional gender roles, which they viewed as a key reason for these men seeking Thai partners over `Western women', who they considered more `independent'. The women thus often presented themselves to a potential \emph{farang} partner in ways that highlighted their ability to fulfil caregiving and household responsibilities.

\subsubsection{Presenting Oneself to \emph{Farangs} Online}\label{sec:presenting-online} 

The participants in our study felt they had no choice but to accept being presented in highly sexualised ways online by bar owners to attract customers (cf.~\Cref{sec:Marketing_Women}). However, they also spoke about how they adopted similar practices of online presentation to attract a \emph{farang}. For example, they exemplified that they would often send photos to \emph{farang} men demonstrating that they were learning English or spending time with Westerners (such as the teachers at the centre). During the fieldwork, the Western volunteers at the centre recalled how the women regularly wanted to get photos with them during or after class to share with \emph{farang} men. The women explained that they shared these pictures because they wanted to convey the message that they would not only be able to communicate in the \emph{farang's} language but also demonstrate that they regularly spent time with \emph{farangs} and understood Western cultural expectations. Many of the shared photos were cross-posted to the women's Facebook pages, where their profiles resembled personal \emph{adverts}. 

Although women used their personal pages differently from the bars and had more control over how they presented themselves, the main goal remained to attract attention from a \emph{farang} man. For example, one woman at the centre took a photo with the Western English teacher. She showed the fieldworker how she sent the image, via Facebook Messenger, to several \emph{farang} men she was talking to, hoping that \emph{farang} men would appreciate her effort to learn English. 

Participants perceived \emph{farang} men as desiring some aspects of Thai women, such as domesticity and traditional women roles, while not desiring aspects such as their lives of poverty or difficult manual labour (both in Pattaya and their hometowns). This was a clear consideration for them when they presented themselves online. The women in our study highlighted that they wanted to show that they would not become too much of a burden or be too unfamiliar with Western ways. Therefore, they constructed their Facebook pages to market themselves as easily able to adapt to the perceived lifestyle of \emph{farangs}. Participants would also alter their online appearance to attract a \emph{farang}. E2, a North American expat working in women's centres in Pattaya, noted that women learned how to use AI tools to modify their pictures so that they would appear to have whiter skin and more `Western features'. They believed that depicting themselves with whiter skin showed that they were not poor or from a lower class, something they thought would be unappealing to \emph{farang} men. 

\subsubsection{Engaging \emph{Farangs} Online}\label{sec:protection-physical-harm} 

Given the risks associated with engaging \emph{farang} men when they visited Pattaya for sex tourism, which came with similar risks to working in the sex industry (cf.~\Cref{sec:Insecurities_Pattaya}), many women in our study noted that searching for a \emph{farang} partner online was the safest and most secure way for them to achieve their goals. They therefore leveraged diverse online resources. This included online dating applications such as Tinder and `Western' social media platforms such as Facebook. In particular, Facebook and Facebook Messenger were the most popular platforms among the women in our study. Women explained how they would often connect with \emph{farang} men via specific Pattaya dating or nightlife groups on Facebook. 

Communicating with \emph{farangs} online also made the women feel more confident in their interactions and, as a result, potentially revealing more about themselves than they would have if they had met in person. There were several examples of this in our data. Participants explained how they, when engaging with \emph{farangs} online, would use services like Google Translate, which helped them feel more confident. Communicating through online platforms had other perceived benefits for the women. For example, some participants mentioned allowing themselves to be more extroverted whilst chatting online because they could not see who they were talking to. P9, a 46-year-old woman from Isan looking for a foreign boyfriend, explained: \textit{``Our challenge is [wanting to have] a foreign boyfriend. So I need to talk to lots of people. Messenger is often easier [than talking in person].''} P46 was worried about meeting strangers in person. Talking online became a `shield' from one-on-one interactions until she felt comfortable enough to meet a \emph{farang} in person. Yet, P41, a 60-year-old woman from Bangkok, exemplified how even when meeting \emph{farangs} online, there would be an expectation that the interaction quickly moved offline: 

\begin{quote}
	\textit{``I have just found a German guy [on] a dating website. I met him through a website, but I do massage. So he came over to our parlour, and I gave him a massage [\ldots] So he asked me to come to his room, and he gave me some money and coffee. I did the massage, and I also had to sleep with him. But now, we are boyfriend and girlfriend.''}
\end{quote}

\noindent Similarly, women who went with a customer as a \emph{holiday girlfriend}\footnote{A `holiday girlfriend' is an extension of sex work where a woman is hired to be a temporary girlfriend to a \emph{farang} while he is in Thailand.} would sometimes meet these men online. As a form of protection, the women would often share the man's photo with a friend or family member before meeting him. For example, one participant explained how they had met their `holiday \emph{farang}' online and was \emph{``so scared. I sent a picture of him to our mum and said, `If I go missing, this is who you should look for'.''}

Throughout interviews with participants, it was not uncommon to hear stories of women who had been killed or had gone missing or thought to have been victims of human trafficking. Sometimes, these threats made women leave the sex industry, but these threats still remained if women wanted to go on a date with a potential \emph{farang} partner. One way women overcame these risks was to move (initial) interactions with \emph{farangs} online, affording them a degree of protection. They recognised that while online interactions posed some risks, e.g.~in the form of scams (cf.~\Cref{sec:scammers}), they allowed them to interact with \emph{farangs} without having to do so in person, revealing their location (often their workplace) and, thus, potentially leaving themselves more vulnerable. 

As presented in~\Cref{sec:collective-security}, some women were aware that meeting \emph{farang} men could potentially lead to physical harm, including being murdered or trafficked, and many of them had experiences of being threatened or abused by clients. Engaging a \emph{farang} online first became a mechanism to decrease the physical risks experienced by many women because they felt that they could disengage if they started feeling unsafe or if the conversation was not leading to their desired results, rather than revealing where they worked. 

\subsection{Collective Protection from \emph{Farangs}}\label{sec:collective-security}

\noindent \Cref{sec:attracting-farang-online} showed why women felt they should turn to online platforms. However, while searching for a \emph{farang} partner online was largely an individual effort, the women relied on collective and shared practices to do so securely. Rather than through familial networks of support, which most women in our study no longer had access to (cf.~\Cref{sec:Losing_Face}), these collective practices were learned through knowledge-sharing between the women, through online groups and at the centre. For example, P5, a 28-year-old from Isan, explained: 

\begin{quote}
    \textit{``We [women at the centre] share education but other things as well. So some [women] have experience getting foreigners to date and visas to go abroad, so we will [talk] about that. We can share about these things [at the centre].''}
\end{quote}

\noindent P5 saw her attendance at the centre as a strategic advantage to learning from women with previous experiences with \emph{farangs} and living abroad. Like for other participants, security for her was rooted in learning how to secure a \emph{farang}, and she had navigated online platforms to find a \emph{farang} husband from Germany, whom she had married and planned to move abroad with. Thus, both successful and unsuccessful stories of \emph{farang} relationships from other women contributed to a collective knowledge that the women felt they could rely on. 

\subsubsection{Identifying Scammers}\label{sec:scammers} 

One central example of collective knowledge and security was demonstrated in how the women discussed scams while trying to find a \emph{farang} partner. Many women explained how they had experienced being scammed through fake \emph{farang} accounts. Often, these scams would involve a pretend \emph{farang} asking for money. Despite often knowing about online scams through collective knowledge, women continued to interact with \emph{farangs} online in the hope that they would find a \emph{farang} and, as a form of protection from physical harm (cf.~\Cref{sec:protection-physical-harm}), avoid having to meet many \emph{farangs} in person. Because of risks embedded in these online scams, some women began to learn the patterns of how scammers would operate. For example, they noted how scam accounts would often send many long messages, expressing how much they loved her and how they desired to send her expensive gifts. Then, the scammer would have a problem occur where the woman needed to send him money, such as to pay taxes on an expensive gift or help him with visa fees. P9, a 46-year-old from Isan, provided an example: 

\begin{quote}
    \textit{``There are many different types of scammers [\ldots] [Some] will say `I've flown to see you, I've reached the airport already.' But then, once he arrives, he will make me talk to a staff member, and they will say he is being evicted and that I should send money [\ldots] but he's just trying to get money out of us.''}
\end{quote}

\noindent Because of the prevalence of these scams, women engaged in knowledge sharing to help build a collective understanding of what patterns to look out for to identify a scammer. For example, P47, a 41-year-old woman from Isan, had a Facebook and TikTok page to help other women find dates and also inform women about scammers. During lunch and before classes at the centre, she would share the page with the other women, encouraging them to follow such pages. On the Facebook page, she would post stories of people she knew who had been scammed, while also posting photos of the scammer to alert others to the scammer's identity. She explained that she had decided to open \emph{``the scammer page''} after having searched for a \emph{farang} for two years. It was a private group -- \textit{``a hotline for scammers''} -- to enable women to share their experiences and develop collective protective practices: 

\begin{quote}
	\textit{``[The women] wouldn't know [about common scams] and they would transfer the money [to the scammers]. There are girls who [\ldots] later find out that [the men] are scammers. So they take a photo and post it in the group [describing the scam] so that everyone knows they [the men] are scammers.''}
\end{quote}

\noindent Throughout the fieldwork we also learned about similar groups that women joined on Line\footnote{Line (https://line.me/) is a popular messaging application in Asia.} and Facebook to learn about strategies to avoid being scammed; these groups became places of educating women on patterns of online scams. Similar to gaining knowledge on finding a \emph{farang}, many women joined these groups online or discussed the scams at the centre. These online groups also made women feel a sense of community and belonging by recognising that they were not on their own. For example, P12, a 40-year-old from Isan, discussed an experience where a \emph{farang} she was speaking to online stole her identity and used it to get a loan from a bank. Consequentially, P12 was both reported to the police and also threatened by the scammer that they would kill her children if she spoke out. P12 turned to online resources, stating: \emph{``There is a Line and Facebook group I'm part of where we would share stories about what had happened to us through scams.''} This online support helped P12 feel less isolated in her experiences with \emph{farangs} online. 

\subsubsection{Scanning \emph{Farangs}}\label{sec:scanning} 

Knowledge sharing often helped many women determine if a certain \emph{farang} was a good choice as a partner, developing a collective understanding of \emph{farang} behaviour that signified `good' relationships. In one of the focus groups, women detailed how they had to \textit{``scan foreigners''} before they could choose them as someone to date. During the focus group, women excitedly spoke amongst themselves sharing strategies that they used to find a \emph{``good''} \emph{farang}. P44, a 27-year-old from Isan, explained: 

\begin{quote}
    \textit{``Not all foreign men are rich. So we have to scan them to see. [We mainly] look at their personality [\ldots] It is good if they want to tell you about themselves and get to know you. Then they are less likely to think of you as a holiday girlfriend. Some of them allow you to go to their home country to see their life and who they are. That case is good. It means they have money. But those that don’t show you that might be an issue. They may be scammers.''}
\end{quote}

\noindent P44's quote reflects how the \emph{scanning} process mainly involved figuring out if a \emph{farang's} financial status would be enough for her to pursue the relationship. If a \emph{farang} showed a willingness to spend money on her -- e.g.~by paying for her flight to visit his country -- that was perceived as proof that he would be willing to financially support her, and most likely also her family. \emph{``Scanning men''} also included learning about \emph{farangs} and their income and building on each other's knowledge of specific jobs to determine whether or not to pursue a relationship with specific \emph{farangs}.

Some women wanted to ensure that a \emph{farang} would not only support her, but also her family. Some participants had children that they wanted to migrate abroad with and others had family members they wanted to continue to support. Financial \emph{scanning} became a way to determine if a \emph{farang} was open to supporting her family. P8, a 46-year-old from Isan, expressed an appreciation for her \emph{farang} husband and explained how he had been integral to her providing the support she desired for her mother. Although P8 had received financial support for her family from a \emph{farang}, she also felt that her relationship included genuine support, care and love, and she trusted her husband to help her through difficult life circumstances. Thus, while participants had something to gain from the \emph{farang}, including support for their family, improved quality of life, better workplace opportunities and migration, some participants felt their relationships with \emph{farangs} extended beyond financial support.

\subsubsection{Teaching at The Centre}\label{sec:teaching-centre}

The centre offered technology classes for the women, which included a 10-day computer class and a two-day mobile phone class. Here, women were taught about basic computer (e.g. Word, PowerPoint, Excel) and mobile phone (e.g. settings, maps, contacts, messaging) functions, and tools such as ChatGPT. On the surface, these classes were designed to improve the women's technological skills to help them get a job outside of sex and massage work. Yet, the fieldwork revealed how these classes were less about building technological capabilities and more about enhancing the women's ability to navigate social and relational pressures, grounded in their precarious circumstances. S1, who taught these computer classes, had been married to a German \emph{farang} and moved to Germany in the early 2000s. While she had left it to the \emph{farang} to handle legal documents such as the marriage license, she later learned that the license was fake and was given four weeks to return to Thailand. This experience impacted how S1 perceived \emph{farang} relationships: \emph{``I've not seen a good relationship. Having a [Thai] woman is like owning a computer. Tomorrow, you [the farang] need an updated version. So you go buy a new one.''} S1's experience of a \emph{farang} relationship, which shaped her views of \emph{farang} relations in general, also directly shaped her computer classes. Here, her teaching focused on protective strategies for women to use when navigating online, such as how to block unwanted contacts or how to identify a scam. 

V1, a volunteer who had led one of the technology classes, explained how the staff had instructed her to teach the women ChatGPT, despite the women struggling to do basic functions such as turning devices on/off or writing a sentence in a Word document. In teaching the women how to use ChatGPT she had asked them to \textit{``ask it any question about basic stuff that is public knowledge''.} This led most of the women to ask questions that related to their search for a \emph{farang}, e.g.~\textit{``how to get married quickly?''} as well as \textit{``many questions about [a foreign] visa or about men abroad.''} V1 had been reluctant to introduce the women to ChatGPT, worrying that they would not be able to critically evaluate the answers they received.

In language classes at the centre, staff equipped women with phrases to help them manage conflicts in their relationships, such as: \emph{``I do not understand, can you speak more slowly?''}, \emph{``I am confused, can you repeat yourself?''} and \emph{``I do not like the tone you are using. Can you speak nicer, please?''}. The staff believed that teaching women these phrases would make them more confident in communicating with a \emph{farang}. The centre also hosted more than 40 lawyers from Bangkok one day to provide legal advice on dating and marrying a \emph{farang}. The session included legal implications of moving abroad or staying in Thailand with a \emph{farang}, visa requirements, family law and the woman's rights if they had children together, and receiving pensions from \emph{farangs}. Women were offered one-to-one legal consultations so they could receive advice related to their specific circumstances. 

\section{Discussion}\label{sec:discussion}

\noindent We contribute to prior security scholarship that calls for research that situates security concerns in the settings under study. We do so in~\Cref{sec:situated-security}. In~\Cref{sec:collective-security-everyday}, we build on the notion of collective security and discuss the centrality of this in our findings. In~\Cref{sec:appropriation-tech}, we discuss how participants focused on using `Western technology', while~\Cref{sec:access} considers how the women conceptualised security in terms of access. In~\Cref{sec:future-work}, we set out potential future research directions for digital security researchers.

\subsection{Situated Digital Security in Pattaya}\label{sec:situated-security} 

\noindent The setting we studied brings to the fore distinct digital security challenges, particularly for groups whose very existence is already at risk. The intertwining of technological and social dimensions of security is exemplified in existing security-driven work with people in marginalised and at-risk contexts, as we report in~\Cref{sec:rw-marginalisation}, highlighting how their security needs are not well served by the technologies they rely upon. For example, the idea of collective security underscores the work by the authors of~\cite{CSCW:MccJenTal24}, who showed how marginalised groups in Lebanon developed collaborative practices to secure themselves and each other. Feelings linked to shame and `othering' were also studied in~\cite{USENIX:MccJenTal23} in related work on marginalised groups in Lebanon, while the authors of~\cite{sambasivan2019they} showed how South Asian women developed performative practices to protect their privacy from intimate and familial relations. 

We show how, to the women in our study, digital security was intimately tied to their social reality. A superficial reading of our findings might suggest that because the women in our study lacked technological know-how, they faced distinct security risks as a result. However, we posit that this would be an incorrect reading. The women in our study learned how to use specific technology for what they considered security to them (cf.~\Cref{sec:scammers} and~\Cref{sec:scanning}); thus, developing particular -- and often elaborate -- practices that provided them with security. Further, as we show, the women relied on digital technology to, for example, search for and attract a \emph{farang}, seek legal advice, identify threats and support each other through collective practices, while they \emph{chose} the technologies, e.g.~Western, they considered could serve their needs. Thus, they developed the technological skills and chose the technologies they considered beneficial for their particular situation. 

As we highlight in \Cref{sec:presenting-online}, many women also uploaded pictures of themselves with the volunteers at the centre, while they curated their social media profiles to align with what they considered attractive to Westerners. The women's adoption of technologies such as generative AI and image-alteration software supported their security goal of attracting a \emph{farang} rather than presenting as a security risk to them. Centre staff, whose lived experiences resembled those of the women, understood such needs and focused their technology classes on ensuring that the women could benefit from having technological access  (cf.~\Cref{sec:teaching-centre}). 

Findings from our study thus point to tensions between the specific context of the women and their understanding of `Western ideals', as expressed through the strategies they (and bar owners, who marketed the women online, cf.~\Cref{sec:Marketing_Women}) adopted to appeal to \emph{farang} men. This influenced the choices the women made about the technologies they chose to rely on for their security -- technologies they considered distinctly `Western': Facebook in particular. This presents a challenge to digital security research; designing \emph{for} groups in distinct marginalised contexts requires a design approach situated in the specific security goals of the specific groups. Specifically, we suggest working towards particular security choices and controls for distinct groups, rooted in their situated context and building on their existing protection strategies. This would also require (re)considering what is considered digital security -- and to whom. 

\subsection{Collective Security in Pattaya}\label{sec:collective-security-everyday}
\noindent Women in Pattaya established collective, trusted networks to search for a \emph{farang} partner and to protect themselves during this process (cf.~\Cref{sec:collective-security}). These networks, formed both in person and online, were essential for navigating the insecurities they faced daily. Rather than emerging in isolation, these networks drew on shared routines, experiences and informal advice, allowing the women to develop collective approaches to their individual (in)security. Our findings reveal a tension between the individual security needs the women sought by looking for a \emph{farang} partner, exposing them to individual risks, and the collective practices adopted to mitigate those risks. These informal systems enabled women to build trust and, by extension, security in the absence of reliable familial or societal support.

Most women migrated to Pattaya from rural communities, leaving their families behind in search of a better future, and took up work in the sex and massage industries. This often left them feeling isolated in an unknown, foreign place. Many of the relationships among women in Pattaya were fraught with conflict and mistrust as they competed for clients and faced workplace bullying (cf.~\Cref{sec:Losing_Face}). The fear of \emph{losing face} also left most women without family support, increasing their reliance on informal networks they formed with other women through the centre and online groups. In particular, through platforms like Facebook and TikTok, the women exchanged information about identifying trustworthy \emph{farang} partners (cf.~\Cref{sec:scanning}), avoiding scams and navigating potential abuse (cf.~\Cref{sec:scammers}). This transformed individual experiences into a collective digital resource where `warning signs' and alerts about unsafe encounters were shared and discussed. 

Our findings reveal how women developed collective protective strategies in their search for security from a \emph{farang}. The importance of finding a ``good'' \emph{farang}, someone who would provide financial security for the woman and her family along with pathways to marriage and migration, became essential for participants' security goals. For instance, the \emph{scanning} of men aimed to determine whether the \emph{farang} saw the woman as a holiday girlfriend or a long-term partner. This practice was a collective one between women who shared similar security goals. Additionally, through knowledge sharing, women learned about the experiences of others, which helped them develop shared protective practices (cf.~\Cref{sec:collective-security}). For example, women who had faced abuse with a \emph{farang} abroad informed others about what to watch out for, questions to ask before marriage and how to seek help when abroad. The \emph{scanning} of men thus functioned as a form of \emph{verification practice} where the women would aim to collectively `verify' the intentions of a \emph{farang}.

We are not the first to highlight the role of informal sharing networks in marginalised contexts. As the authors of~\cite{USENIX:MccJenTal23} note in their research on post-conflict Lebanon, informal networks, including digital ones, often protect those lacking support from state or formal institutions. The women in our study enhanced their security by forming practical security networks through their shared, everyday practices.

We are also not the first to highlight the protective benefits for sex workers in using online and collective resources instead of in-person methods to find potential customers (cf.~\Cref{sec:setting}). As noted in~\cite{CHI:StrClaLai19,CHI:StrLaiCom17}, organisations focusing on sex workers' rights leverage digital technology to cater to sex workers through the sharing of information through online networks. Our research contributes to this literature by examining the role of the centre in the women's (digital) security. As we highlight in~\Cref{sec:teaching-centre}, technology classes were taught from the women's grounded perspectives, focusing on their specific security needs and goals. However, as we show in~\Cref{sec:collective-security} the women in our study also formed their own networks to access tailored information to meet their needs. For example, while the centre provided technology classes on online protections, such as blocking certain contacts, the women relied on informal (online) networks for specific and personal support and advice; examples include advice on appealing to a \emph{farang} partner (cf.~\Cref{sec:presenting-online}) and how to avoid scammers (cf.~\Cref{sec:scammers}). These informal networks provided an additional `layer' of security, which the centre did not provide, as the staff often did not teach specific scamming techniques or the evolving risks associated with finding a \emph{farang} partner. 

More broadly, existing work has shown how the transition to online sex work during and after the COVID-19 pandemic has brought certain benefits to sex workers, e.g.~\cite{CSCW:HamBarRed22}. For many, the shift to online sex work is seen to enhance sex workers' agency and security by reducing risks associated with in-person sex work~\cite{barwulor2021disadvantaged,castle2008ordering, sanders2016our,scoular2019beyond,veena2007revisiting}. While our work did not specifically focus on sex work, we found that by adopting online strategies to find a \emph{farang} partner, the women could protect themselves from some risks associated with sex work, including violent and abusive clients, STDs, trafficking and the risk of being murdered. However, as our findings also show, while the women leveraged online platforms to find \emph{farangs}, the interaction quickly transitioned to in-person meetings where the women would be expected to `perform' certain services (cf.~\Cref{sec:protection-physical-harm}). Thus, while the women expressed feeling safer searching for and initiating conversations with \emph{farangs} online, they also exemplified how this environment presented new risks such as being scammed or trafficked, while the protection the online environment afforded them was temporal.

\subsection{Perceived Security of and in `Western Tech'}\label{sec:appropriation-tech}

\noindent The participants in our study voiced different insecurities, most of which were founded in the social and economic conditions that shaped their daily lives -- and struggles. While their security concerns were predominantly tied to their lack of financial means and familial bonds, they often appeared in the form of technological insecurity. The women's continuous search for a \emph{farang}, with the hope of financially securing themselves and their families, exposed the women to distinct security risks. These included online scams on dating platforms (cf.~\Cref{sec:scammers}), the online marketing of women by bar owners or \emph{mamasans} (cf.~\Cref{sec:Marketing_Women}), while the women tried to mitigate against such risks through knowledge sharing both at the centre and through dedicated online groups (cf.~\Cref{sec:collective-security}). The women built technological capabilities that served their specific goals. This was seen in how they devised distinct security practices, including: 

\begin{itemize}
	\setlength \itemsep{0.2em}
	\item collectively identifying scammers and sharing their `patterns of operation' through online groups;
	\item `scanning' \emph{farangs} online to protect themselves from potentially entering abusive and less financially appealing relationships;
	\item using image alteration technology to make their skin appear whiter (and with more `Western features') in an attempt to, in their view, make themselves more appealing to \emph{farangs}; and
	\item using generative AI tools to retrieve information about relationships, how to get a visa and what (\emph{farang}) men would find attractive.
\end{itemize}

\noindent Thus, while digital security was not a goal in itself for the women in our study, they relied on technology in their search for a \emph{farang} (e.g.~by presenting themselves online, cf.~\Cref{sec:presenting-online}). This reliance was rooted in different forms of security, including minimising physical abuse from clients (cf.~\Cref{sec:protection-physical-harm}) and financial security. In other words, while participants demonstrated a reliance on technology for their security, this reliance was not influenced by the security of a given technology, even when centre staff raised concerns about their use of a certain technology (cf.~\Cref{sec:teaching-centre}). An important reminder here is that the women in our study were not unaware of the security risks they faced when engaging with \emph{farangs} online (cf.~\Cref{sec:Benefits_Farang}). However, for them, this was often considered a safer option than trying to attract a \emph{farang} in their workplace, i.e.~in bars or in the street, where they described several experiences of physical security risks -- both their own and the risks experienced by others. 

\subsection{Access Benefit, not Access Control}\label{sec:access}

\noindent Computer and technology classes at the centre were predominantly taught by staff members who had themselves experiences that resembled those of the women in our study. This included experiences of sex work, marrying a \emph{farang} and migrating abroad with him, as well as experiences of scams, threats and abuse (cf.~\Cref{sec:teaching-centre}). Thus, while the women were taught basic technological skills in such classes, the skills they learned were situated in their social reality. Through these classes the women gained \emph{access} to different technological resources, where the teaching focused on providing knowledge that enabled the women to access the benefits of this resource -- rather than controlling access to it. In their \emph{theory of access}, Ribot and Peluso conceptualise access as the ability to realise benefit from something~\cite{RS:RibPel03}, where access includes the ability to benefit from people, communities and social, political and economic relationships.\footnote{This theory of access has been used to understand what access to digital technology means in the context of other underserved groups (see e.g.~\cite{CHI:colesJensen2019}).} We see that for the women in our study, their main goal of learning technological skills was to gain access to the benefit of a technology (e.g.~enabling them to search for a \emph{farang}), not controlling access to that benefit. The teachers' lived experiences enabled this learning. 

Our findings further epitomise the significance of this. We show how Western volunteers, who had been instructed by centre staff to teach the women how to use ChatGPT, were reluctant to do so, citing concerns over their lack of skills to detect misinformation (cf.~\Cref{sec:teaching-centre}). With no prior experiences that resembled those of the women to draw upon, the volunteers approached the teaching by focusing on the risks inherent in the technology and, thus, wanting to control the women's access to it. In contrast, centre staff educated women on how to use digital technology, but with a specific focus on how to navigate online interactions with a \emph{farang}, for example. Technology classes provided by the centre were designed from the perspective of women in Pattaya, not from the technological affordabilities or security features that they might gain from downloading a given secure messaging application or learning how to set a particularly strong password on their phone. We show that for the centre, computer education was focused on access benefit, not access control, supporting the women in building capabilities to live the life they could realistically achieve.

Our work thus also speaks to the role of security educators and trainers in `at-risk' contexts more generally, highlighting the need for security learning to be rooted in the social experiences and contexts as much as in the technology. This is also exemplified in broader work, e.g.~research with security trainers in activist settings shows the importance of them being situated in the community they serve~\cite{EWUS:ErmHalMus17}. 

\subsection{Implications and Future Work}\label{sec:future-work}

\noindent Through immersion in the setting under study, our work uncovered what security meant and how it was practised by \emph{these} women, in \emph{this} setting. We consider this to be the most fundamental contribution of our work. Thus, our findings (cf.~\Cref{sec:findings}), enabled through extended fieldwork, shed light on the social undercurrents that shaped technology and security choices for the women in our study. From this position, we concretise the implications of our work for digital security research. 

\subsubsection{Ethnography} We consider our methodology a direct contribution to future security and privacy research. We demonstrate how an immersive approach in a specific, hard-to-access, non-WEIRD (cf.~\cite{CHI:LSBCOR21,USENIX:HasDaiAki24}) setting, uncovers aspects of technology use and related security practices that are often not interrogated and thus poorly understood in security and privacy research. At the same time, we see a greater focus on designing for such settings within security and privacy, which requires a diversity of methods (cf.~\cite{USENIX:MccJenTal23,USENIX:SCBAPB22}). We demonstrate one such approach: ethnography. Our ethnographic approach enabled us to engage with the setting under study through a ground-up and longer-term programme of research. Through fieldwork we uncovered how digital security for the women in our study was intertwined with a series of broader insecurities. While ethnographic approaches underlie some recent security and privacy research (e.g.~\cite{USENIX:MccJenTal23,CSCW:MccJenTal24,CHI:SNBMAH24}), much (qualitative) security and privacy research is often conducted \emph{ex situ} as also noted in~\cite{USENIX:BroJenAlb25}. We posit, however, that it is through immersive methods applied \emph{in situ} that we can learn that which is otherwise `hidden' and not articulated in, say, interviews or focus groups. For example, in~\cite{CHI:KKSC16} the authors use the idea of ``undercurrents'' to understand how security is considered, decided upon and practised in social movements. While our object of study differs, we echo the need for security researchers to engage with and understand the social foundations of security to contextualise technology and design decisions. 

Our ethnographic fieldwork was limited to one centre in Pattaya. Future work might consider conducting immersive research in a diversity of centres and wider settings, expanding our work and the work of~\cite{CHI:SNBMAH24}, for example, to broaden and deepen the digital security insights uncovered in these works.

\subsubsection{Situated Security} Situating digital security in the setting we studied moves beyond advocating for specific technological interventions. Rather, we call for a situated design approach that prioritises the conditions that shape people's lived security needs (cf.~\Cref{sec:situated-security}). Such a design approach acknowledges the everyday -- the mundane -- security needs of distinct groups while, more abstractly, considering how social and economic conditions shape digital security for people in marginalised contexts (cf.~\Cref{sec:appropriation-tech}). This diverges from universal or `Western' design paradigms, but rests on collaborative and participatory approaches. From this grounding, we point to (re)considering security controls and choices where security is enabled through the availability and accessibility of a technology, e.g.~our finding on access benefit or access control (cf.~\Cref{sec:access}). 

Our work uncovered distinct security threats facing the women in our study, including being scammed by \emph{farang} men and experiences of technology-facilitated intimate abuse. While our work does not focus on these threats, we consider this a pressing direction for future security and privacy research, with most existing research on sex work and technology abuse situated in Global North contexts (cf.~\Cref{sec:setting}).

\subsubsection{Collaborative} Our work points to how the women in our study relied on shared knowledge to protect themselves, often realised during their (brief) time at the centre. Thus, future security and privacy research might build on collaborative and participatory design practices in such safe spaces~\cite{AkaYee24} to develop community-driven, technology-facilitated secure sharing networks. Our study brought to light the importance of collaborative and informal sharing networks, which functioned as vital security infrastructures for the women in our study (cf.~\Cref{sec:collective-security-everyday}). Future design work might consider implementing methods for trusted information sharing, enabling groups to collaboratively identify threats and develop verification practices as they materialise in \emph{their} context and daily encounters. 

\subsubsection{Education} We show how the role of technology and security educators extends beyond considering the security of a technology in this setting, cf.~our finding that the women in our study appropriated `Western technology' to serve their needs (\Cref{sec:appropriation-tech}), which suggests digital security education that is designed with the community and for their context. This diverges from much security education for at-risk communities which is often delivered by external experts and/or organisations. Further, in~\Cref{sec:access}, we highlight how enabling access to benefits was considered more important than enforcing strict access controls. While this might not chime with expert security advice, we consider this an important implication of our work: to be effective, digital security education and training needs to be rooted in people's lived experiences; enabling people to be supported to utilise technology for their specific aims, such as financial stability. 

\section{Conclusion}
\noindent Our work, enabled through our ethnographic study, uncovered how Thai women's search for a \emph{farang} partner exposed them to distinct insecurities that they tried to protect against through collective practices and the sharing of knowledge. In setting out directions for future research, we point to the need to situate digital security research in the settings being considered and designed for.  

\section{Acknowledgments}

\noindent This work would not exist without the many contributions from the women at the centre in Pattaya, who so generously gave of their time to speak to us and shared their lived experiences. We thank Katie Willis for helpful discussions in relation to the development of this work. The research of Robinson was supported by the UKRI as part of the Centre for Doctoral Training in Cyber Security for the Everyday at Royal Holloway, University of London (EP/S021817/1). We would also like to acknowledge the National Research Council of Thailand for allowing us to conduct this research in Thailand. 

\bibliographystyle{plain}
\bibliography{local}

\begin{thebibliography}{10}

\bibitem{AkaYee24}
Yoko Akama and Joyce Yee.
\newblock {\em Entanglements of designing social innovation in the
  Asia-Pacific}.
\newblock Routledge, 2024.

\bibitem{USENIX:AlbFle23}
Wael Albayaydh and Ivan Flechais.
\newblock Examining {P}ower {D}ynamics and {U}ser {P}rivacy in {S}mart
  {T}echnology use among {J}ordanian {H}ouseholds.
\newblock In {\em 32nd USENIX Security Symposium (USENIX Security 23)}, pages
  4643--4659, 2023.

\bibitem{USENIX:ABJM21}
Martin~R Albrecht, Jorge Blasco, Rikke~Bjerg Jensen, and Lenka Marekov{\'a}.
\newblock Collective {I}nformation {S}ecurity in {L}arge-{S}cale {U}rban
  {P}rotests: the {C}ase of {H}ong {K}ong.
\newblock In {\em 30th USENIX Security Symposium}, pages 3363--3380, 2021.

\bibitem{CSCW:ArmTalVla22}
Sarah Armouch, Reem Talhouk, and Vasilis Vlachokyriakos.
\newblock {Revolting from abroad: The formation of a Lebanese transnational
  public}.
\newblock {\em Proceedings of the ACM on Human-Computer Interaction},
  6(CSCW2):1--28, 2022.

\bibitem{barwulor2021disadvantaged}
Catherine Barwulor, Allison McDonald, Eszter Hargittai, and Elissa~M Redmiles.
\newblock “{D}isadvantaged in the {A}merican-dominated {I}nternet”: Sex,
  {W}ork, and {T}echnology.
\newblock In {\em Proceedings of the 2021 CHI Conference on Human Factors in
  Computing Systems}, pages 1--16, 2021.

\bibitem{bishop2024bar}
Simon Bishop and Onn Laingoen.
\newblock From the {B}ar to the {C}owshed: the {I}mpact of {COVID-19} on
  {F}emale {S}ex {W}orkers in {P}attaya, {T}hailand.
\newblock {\em Culture, Health \& Sexuality}, 26(12):1543--1555, 2024.

\bibitem{QRP:braun2021}
Virginia Braun and Victoria Clarke.
\newblock One {S}ize {F}its {A}ll? what {C}ounts as {Q}uality {P}ractice in
  ({R}eflexive) {T}hematic {A}nalysis?
\newblock {\em Qualitative {R}esearch in {P}sychology}, 18(3):328--352, 2021.

\bibitem{QESEH:BraCla21}
Virginia Braun and Victoria Clarke.
\newblock To {S}aturate or not to {S}aturate? questioning {D}ata {S}aturation
  as a {U}seful {C}oncept for {T}hematic {A}nalysis and {S}ample-size
  {R}ationales.
\newblock {\em Qualitative {R}esearch in {S}port, {E}xercise and {H}ealth},
  13(2):201--216, 2021.

\bibitem{braun2021saturate}
Virginia Braun and Victoria Clarke.
\newblock To saturate or not to saturate? questioning data saturation as a
  useful concept for thematic analysis and sample-size rationales.
\newblock {\em Qualitative research in sport, exercise and health},
  13(2):201--216, 2021.

\bibitem{braun2022starting}
Virginia Braun, Victoria Clarke, and Nikki Hayfield.
\newblock ‘{A} {S}tarting {P}oint for {Y}our {J}ourney, {N}ot a {M}ap’:
  {N}ikki {H}ayfield in {C}onversation with {V}irginia {B}raun and {V}ictoria
  {C}larke {A}bout {T}hematic {A}nalysis.
\newblock {\em Qualitative Research in Psychology}, 19(2):424--445, 2022.

\bibitem{RH:braun2016}
Virginia Braun, Victoria Clarke, and Paul Weate.
\newblock Using {T}hematic {A}nalysis in {S}port and {E}xercise {R}esearch.
\newblock In {\em Routledge {H}andbook of {Q}ualitative {R}esearch in {S}port
  and {E}xercise}, pages 213--227. Routledge, 2016.

\bibitem{USENIX:BroJenAlb25}
Mikaela Brough, Rikke~Bjerg Jensen, and Martin~R Albrecht.
\newblock {On the Virtues of Information Security in the UK Climate Movement}.
\newblock In {\em 34th USENIX Security Symposium}, pages 5091--5110, 2025.

\bibitem{cameron2005migration}
Michael~Patrick Cameron and Steven Lim.
\newblock Migration, {H}ousehold {C}omposition and {C}hild {W}elfare in {R}ural
  {N}ortheast {T}hailand.
\newblock {\em Department of Economics Working Paper Series}, 2005.

\bibitem{castle2008ordering}
Tammy Castle and Jenifer Lee.
\newblock Ordering {S}ex in {C}yberspace: a {C}ontent {A}nalysis of {E}scort
  {W}ebsites.
\newblock {\em International Journal of Cultural Studies}, 11(1):107--121,
  2008.

\bibitem{chiablaem:2020aNALYSIS}
Parichat Chiablaem.
\newblock A {N}eeds {A}nalysis of {T}hai {M}assage {E}mployees'({T}hai
  {M}assage {T}herapists, {R}eceptionists, and {M}anagers) {E}nglish
  {C}ommunication {S}kills in the {C}ity of {P}attaya, {C}honburi.
\newblock {\em rEFLections}, 27(2):274--292, 2020.

\bibitem{CHI:colesJensen2019}
Lizzie Coles-Kemp and Rikke~Bjerg Jensen.
\newblock Accessing a {N}ew {L}and: {D}esigning for a {S}ocial
  {C}onceptualisation of {A}ccess.
\newblock In {\em Proceedings of the 2019 CHI Conference on Human Factors in
  Computing Systems}, pages 1--12, 2019.

\bibitem{CHI:ColJenTal18}
Lizzie Coles-Kemp, Rikke~Bjerg Jensen, and Reem Talhouk.
\newblock In a {N}ew {L}and: {M}obile {P}hones, {A}mplified {P}ressures and
  {R}educed {C}apabilities.
\newblock In {\em Proceedings of the 2018 CHI Conference on Human Factors in
  Computing Systems}, pages 1--13, 2018.

\bibitem{Craig2024-ae}
Bella Craig.
\newblock Sex {W}orkers in {P}eril.
\newblock
  \url{https://www.bangkokpost.com/thailand/special-reports/2803754/sex-workers-in-peril},
  June 2024.
\newblock Accessed: 2025-5-20.

\bibitem{crawford2016mapping}
Adam Crawford and Steven Hutchinson.
\newblock {M}apping the {C}ontours of ‘{E}veryday {S}ecurity’: {T}ime,
  {S}pace and {E}motion.
\newblock {\em British {J}ournal of {C}riminology}, 56(6):1184--1202, 2016.

\bibitem{Duangdee2020-ke}
Vijitra Duangdee.
\newblock Coronavirus: {I}n {T}hailand, {S}ex {W}ork {M}oves {O}nline as
  {P}andemic {B}atters {B}usiness.
\newblock
  \url{https://www.scmp.com/week-asia/lifestyle-culture/article/3085968/no-go-bars-thailand-sex-work-moves-online-pandemic},
  May 2020.
\newblock Accessed: 2025-5-20.

\bibitem{dupuis1998home}
Ann Dupuis and David~C Thorns.
\newblock Home, home ownership and the search for ontological security.
\newblock {\em The sociological review}, 46(1):24--47, 1998.

\bibitem{WEF:emerson2011}
Robert~M Emerson, Rachel~I Fretz, and Linda~L Shaw.
\newblock {\em Writing {E}thnographic {F}ieldnotes}.
\newblock University of Chicago press, 2011.

\bibitem{EWUS:ErmHalMus17}
Ksenia Ermoshina, Harry Halpin, and Francesca Musiani.
\newblock Can {J}ohnny {B}uild a {P}rotocol? {C}o-ordinating {D}eveloper and
  {U}ser {I}ntentions for {P}rivacy-enhanced {S}ecure {M}essaging {P}rotocols.
\newblock In {\em European Workshop on Usable Security}, pages 1--13, 2017.

\bibitem{eschle2018nuclear}
Catherine Eschle.
\newblock Nuclear (in) {S}ecurity in the {E}veryday: {P}eace {C}ampers as
  {E}veryday {S}ecurity {P}ractitioners.
\newblock {\em Security Dialogue}, 49(4):289--305, 2018.

\bibitem{PRD:Pattaya2024}
The Government Public Relations~Department Foreign~Office.
\newblock Pattaya {A}ims for 27 {M}illion {T}ourists in 2024.
\newblock {\em The Government Public Relations Department}, 2024.

\bibitem{CSCW:FHTGCRD19}
Diana Freed, Sam Havron, Emily Tseng, Andrea Gallardo, Rahul Chatterjee, Thomas
  Ristenpart, and Nicola Dell.
\newblock ``is {M}y {P}hone {H}acked?'' {A}nalyzing {C}linical {C}omputer
  {S}ecurity {I}nterventions with {S}urvivors of {I}ntimate {P}artner
  {V}iolence.
\newblock {\em Proceedings of the ACM on Human-Computer Interaction},
  3(CSCW):1--24, 2019.

\bibitem{CSCW:FPMLRD17}
Diana Freed, Jackeline Palmer, Diana~Elizabeth Minchala, Karen Levy, Thomas
  Ristenpart, and Nicola Dell.
\newblock Digital {T}echnologies and {I}ntimate {P}artner {V}iolence: A
  {Q}ualitative {A}nalysis with {M}ultiple {S}takeholders.
\newblock {\em Proceedings of the ACM on human-computer interaction},
  1(CSCW):1--22, 2017.

\bibitem{USENIX:GHRR22}
Christine Geeng, Mike Harris, Elissa Redmiles, and Franziska Roesner.
\newblock ``{L}ike {L}esbians {W}alking the {P}erimeter'': {E}xperiences of
  {US} {LGBTQ+} {F}olks with {O}nline {S}ecurity, {S}afety, and {P}rivacy
  {A}dvice.
\newblock In {\em 31st USENIX Security Symposium}, 2022.

\bibitem{giddens2016modernity}
Anthony Giddens.
\newblock Modernity and self-identity.
\newblock In {\em Social theory re-wired}, pages 512--521. Routledge, 2016.

\bibitem{gunning2022you}
Jeroen Gunning and Dima Smaira.
\newblock Who you {G}onna call? {T}heorising {E}veryday {S}ecurity {P}ractices
  in {U}rban {S}paces with {M}ultiple {S}ecurity {A}ctors--the {C}ase of
  {B}eirut's {S}outhern {S}uburbs.
\newblock {\em Political Geography}, 98:102485, 2022.

\bibitem{CSCW:HamBarRed22}
Vaughn Hamilton, Hanna Barakat, and Elissa~M Redmiles.
\newblock Risk, {R}esilience and {R}eward: {I}mpacts of {S}hifting to {D}igital
  {S}ex {W}ork.
\newblock {\em Proceedings of the ACM on Human-Computer Interaction},
  6(CSCW2):1--37, 2022.

\bibitem{EPP:hammersley2019}
Martyn Hammersley and Paul Atkinson.
\newblock {\em Ethnography: Principles in practice}.
\newblock Routledge, 2019.

\bibitem{hasayotin2024empowerment}
Khwanchol Hasayotin, Rattanavalee Maisak, Rattanawadee Setthajit, Thanaphon
  Ratchatakulpat, Wannapa Naburana, and Adui Supanut.
\newblock Empowerment of {SMEs} and {E}ntrepreneurial {E}cosystems: a
  {Q}ualitative {S}tudy on {D}iversifying {P}attaya's {E}conomy.
\newblock {\em Revista de Gest{\~a}o Social e Ambiental}, 18(7):1--30, 2024.

\bibitem{USENIX:HasDaiAki24}
Ayako~A Hasegawa, Daisuke Inoue, and Mitsuaki Akiyama.
\newblock How {WEIRD} is {U}sable {P}rivacy and {S}ecurity {R}esearch?
\newblock In {\em 33rd USENIX Security Symposium}, 2024.

\bibitem{higate2010space}
Paul Higate and Marsha Henry.
\newblock Space, {P}erformance and{E}everyday {S}ecurity in the {P}eacekeeping
  {C}ontext.
\newblock {\em International Peacekeeping}, 17(1):32--48, 2010.

\bibitem{Hung2024-us}
Jason Hung.
\newblock Southeast {A}sia must {C}rack {D}own on {O}nline {S}ex {W}ork.
\newblock
  \url{https://asia.nikkei.com/Opinion/Southeast-Asia-must-crack-down-on-online-sex-work},
  October 2024.
\newblock Accessed: 2025-5-20.

\bibitem{huysmans2006politics}
Jef Huysmans.
\newblock {\em The {P}olitics of {I}nsecurity: {F}ear, {M}igration and {A}sylum
  in the {EU}}.
\newblock Routledge, 2006.

\bibitem{CHI:JenColTal20}
Rikke~Bjerg Jensen, Lizzie Coles-Kemp, and Reem Talhouk.
\newblock When the {C}ivic {T}urn {T}urns {D}igital: {D}esigning {S}afe and
  {S}ecure {R}efugee {R}esettlement.
\newblock In {\em Proceedings of the 2020 CHI Conference on Human Factors in
  Computing Systems}, pages 1--14, 2020.

\bibitem{jones2019security}
Peris Jones and Wangui Kimari.
\newblock Security {B}eyond the {M}en: {W}omen and their {E}veryday {S}ecurity
  {A}pparatus in {M}athare, {N}airobi.
\newblock {\em Urban {S}tudies}, 56(9):1835--1849, 2019.

\bibitem{joque2020deconstructing}
Justin Joque and SM~Taiabul Haque.
\newblock Deconstructing {C}ybersecurity: From {O}ntological {S}ecurity to
  {O}ntological {I}nsecurity.
\newblock In {\em Proceedings of the New Security Paradigms Workshop 2020},
  pages 99--110, 2020.

\bibitem{CHI:KKSC16}
Yong~Ming Kow, Yubo Kou, Bryan Semaan, and Waikuen Cheng.
\newblock {Mediating the Undercurrents: Using Social Media to Sustain a Social
  Movement}.
\newblock In {\em Proceedings of the 2016 CHI Conference on Human Factors in
  Computing Systems}, CHI '16, page 3883–3894, New York, NY, USA, 2016.
  Association for Computing Machinery.

\bibitem{JID:le2015}
Ann Le~Mare, Buapun Promphaking, and Jonathan Rigg.
\newblock Returning {H}ome: The {M}iddle-income {T}rap and {G}endered {N}orms
  in {T}hailand.
\newblock {\em Journal of International Development}, 27(2):285--306, 2015.

\bibitem{CHI:LHKZH20}
Ada Lerner, Helen~Yuxun He, Anna Kawakami, Silvia~Catherine Zeamer, and Roberto
  Hoyle.
\newblock Privacy and {A}ctivism in the {T}ransgender {C}ommunity.
\newblock In {\em Proceedings of the 2020 CHI Conference on Human Factors in
  Computing Systems}, pages 1--13, 2020.

\bibitem{CHI:LSBCOR21}
Sebastian Linxen, Christian Sturm, Florian Br{\"u}hlmann, Vincent Cassau, Klaus
  Opwis, and Katharina Reinecke.
\newblock How {W}eird is {CHI?}
\newblock In {\em Proceedings of the 2021 {CHI} {C}onference on {H}uman
  {F}actors in {C}omputing {S}ystems}, pages 1--14, 2021.

\bibitem{DMM:longjit2013}
Chootima Longjit and Douglas~G Pearce.
\newblock Managing a {M}ature {C}oastal {D}estination: {P}attaya, {T}hailand.
\newblock {\em Journal of Destination Marketing \& Management}, 2(3):165--175,
  2013.

\bibitem{CHI:MOTSWSMCC17}
Tara Matthews, Kathleen O'Leary, Anna Turner, Manya Sleeper, Jill~Palzkill
  Woelfer, Martin Shelton, Cori Manthorne, Elizabeth~F Churchill, and Sunny
  Consolvo.
\newblock Stories from survivors: Privacy \& security practices when coping
  with intimate partner abuse.
\newblock In {\em Proceedings of the 2017 CHI conference on human factors in
  computing systems}, pages 2189--2201, 2017.

\bibitem{USENIX:MccJenTal23}
Jessica McClearn, Rikke~Bjerg Jensen, and Reem Talhouk.
\newblock Othered, {S}ilenced and {S}capegoated: Understanding the {S}ituated
  {S}ecurity of {M}arginalised {P}opulations in {L}ebanon.
\newblock In {\em 32nd USENIX Security Symposium (USENIX Security 23)}, pages
  4625--4642, 2023.

\bibitem{CSCW:MccJenTal24}
Jessica McClearn, Rikke~Bjerg Jensen, and Reem Talhouk.
\newblock Security {P}atchworking in {L}ebanon: {I}nfrastructuring {A}cross
  {F}ailing {I}nfrastructures.
\newblock {\em Proceedings of the ACM on Human-Computer Interaction},
  8(CSCW1):1--26, 2024.

\bibitem{mcdonald2021s}
Allison McDonald, Catherine Barwulor, Michelle~L Mazurek, Florian Schaub, and
  Elissa~M Redmiles.
\newblock ``it's {S}tressful {H}aving all these {P}hones": {I}nvestigating
  {S}ex {W}orkers' {S}afety {G}oals, {R}isks, and {P}ractices {O}nline.
\newblock In {\em 30th USENIX Security Symposium (USENIX Security 21)}, pages
  375--392, 2021.

\bibitem{ijir:McKenzie2021}
Jessica McKenzie and Kajai~C Xiong.
\newblock Fated for {F}oreigners: Ecological {R}ealities {S}hape {P}erspectives
  of {T}ransnational {M}arriage in {N}orthern {T}hailand.
\newblock {\em International Journal of Intercultural Relations}, 82:121--134,
  2021.

\bibitem{SII:mcsweeney1999}
Bill McSweeney.
\newblock {\em Security, {I}dentity and {I}nterests: a {S}ociology of
  {I}nternational {R}elations}.
\newblock Number~69. Cambridge University Press, 1999.

\bibitem{ALQ:muir2023}
Rebecca Muir.
\newblock From {D}ata to {I}nsights: Developing a {T}ool to {E}nhance our
  {D}ecision {M}aking using {R}eflexive {T}hematic {A}nalysis and {Q}ualitative
  {E}vidence.
\newblock {\em Journal of the Australian Library and Information Association},
  72(2):150--165, 2023.

\bibitem{Nguyen2025-no}
Binh Nguyen.
\newblock Man {J}ailed 15 {Y}ears for {B}rokering {O}nline {S}ex {W}ork
  {I}nvolving {T}housands of {W}omen.
\newblock
  \url{https://e.vnexpress.net/news/news/crime/man-jailed-15-years-for-brokering-online-sex-work-involving-thousands-of-women-4873903.html},
  April 2025.
\newblock Accessed: 2025-5-20.

\bibitem{nyman2021everyday}
Jonna Nyman.
\newblock The {E}veryday {L}ife of {S}ecurity: Capturing {S}pace, {P}ractice,
  and {A}ffect.
\newblock {\em International Political Sociology}, 15(3):313--337, 2021.

\bibitem{JAH:Phannarat:2016}
Jirawitt Phannarat, Souneth Phothisane, and Homhuan Buarapha.
\newblock Channels to {M}eet {F}oreign {P}artners and the {D}aily {L}ife
  {A}daptation in the {C}urrent {S}ociety of the {S}outhern {T}hai {F}amilies
  whose {M}embers are {M}arried to {F}oreigners.
\newblock {\em Journal of Arts and Humanities}, 5(8):50--58, 2016.

\bibitem{phipps2017routines}
Marcus Phipps and Julie~L Ozanne.
\newblock Routines disrupted: Reestablishing security through practice
  alignment.
\newblock {\em Journal of Consumer Research}, 44(2):361--380, 2017.

\bibitem{SD:plambech2023}
Sine Plambech.
\newblock ‘{M}y {B}ody is my {P}iece of {L}and’: {I}ndebted {D}eportation
  {A}mong {U}ndocumented {M}igrant {S}ex {W}orkers from {T}hailand and
  {N}igeria in {E}urope.
\newblock {\em Security Dialogue}, 54(6):586--601, 2023.

\bibitem{APJMR:Pomsema2015}
Chantaya Pomsema, Boonsom Yodmalee, and Sastra Lao-Akka.
\newblock Foreigners’ {W}ives: {C}ross-cultural {M}arriage of {R}ural {T}hai
  {W}omen in {I}san, {T}hailand.
\newblock {\em Asia Pacific Journal of Multidisciplinary Research},
  3(3):11--15, 2015.

\bibitem{RS:RibPel03}
Jesse~C Ribot and Nancy~Lee Peluso.
\newblock A {T}heory of {A}ccess.
\newblock {\em Rural {S}ociology}, 68(2):153--181, 2003.

\bibitem{VPS:roe2008}
Paul Roe.
\newblock The ‘{V}alue’of {P}ositive {S}ecurity.
\newblock {\em {R}eview of {I}nternational {S}tudies}, 34(4):777--794, 2008.

\bibitem{CHI:SNBMAH24}
Pratyasha Saha, Nadira Nowsher, Ayien~Utshob Baidya, Nusrat~Jahan Mim,
  Syed~Ishtiaque Ahmed, and SM~Taiabul Haque.
\newblock Computing and the {S}tigmatized: {T}rust, {S}urveillance, and
  {S}patial {P}olitics with the {S}ex {W}orkers in {B}angladesh.
\newblock In {\em Proceedings of the 2024 CHI Conference on Human Factors in
  Computing Systems}, pages 1--22, 2024.

\bibitem{sambasivan2019they}
Nithya Sambasivan, Amna Batool, Nova Ahmed, Tara Matthews, Kurt Thomas,
  Laura~Sanely Gayt{\'a}n-Lugo, David Nemer, Elie Bursztein, Elizabeth
  Churchill, and Sunny Consolvo.
\newblock ``{T}hey {D}on't {L}eave us {A}lone {A}nywhere we {G}o" {G}ender and
  {D}igital {A}buse in {S}outh {A}sia.
\newblock In {\em {P}roceedings of the 2019 CHI Conference on Human Factors in
  Computing Systems}, pages 1--14, 2019.

\bibitem{sanders2016our}
Teela Sanders, Laura Connelly, and Laura~Jarvis King.
\newblock On our {O}wn {T}erms: {T}he {W}orking {C}onditions of
  {I}nternet-{B}ased {S}ex {W}orkers in the {UK}.
\newblock {\em Sociological {R}esearch {O}nline}, 21(4):133--146, 2016.

\bibitem{scoular2019beyond}
Jane Scoular, Jane Pitcher, Teela Sanders, Rosie Campbell, and Stewart
  Cunningham.
\newblock Beyond the {G}aze and {W}ell beyond {W}olfenden: {T}he {P}ractices
  and {R}ationalities of {R}egulating and {P}olicing {S}ex {W}ork in the
  {D}igital {A}ge.
\newblock {\em Journal of {L}aw and {S}ociety}, 46(2):211--239, 2019.

\bibitem{SIM:Inequality2019}
Eelynn Sim.
\newblock Thailand’s {I}nequality: {U}npacking the {M}yths and {R}eality of
  {I}san.
\newblock {\em The Asia Foundation}, 2019.

\bibitem{SP:SLIRK18}
Lucy Simko, Ada Lerner, Samia Ibtasam, Franziska Roesner, and Tadayoshi Kohno.
\newblock Computer {S}ecurity and {P}rivacy for {R}efugees in the {U}nited
  {S}tates.
\newblock In {\em 2018 IEEE Symposium on Security and Privacy (SP)}, pages
  409--423. IEEE, 2018.

\bibitem{CHI:SMOTWSOSC19}
Manya Sleeper, Tara Matthews, Kathleen O'Leary, Anna Turner, Jill~Palzkill
  Woelfer, Martin Shelton, Andrew Oplinger, Andreas Schou, and Sunny Consolvo.
\newblock Tough times at transitional homeless shelters: Considering the impact
  of financial insecurity on digital security and privacy.
\newblock In {\em Proceedings of the 2019 CHI Conference on Human Factors in
  Computing Systems}, pages 1--12, 2019.

\bibitem{USENIX:SCBAPB22}
Julia S{\l}upska, Selina Cho, Marissa Begonia, Ruba Abu-Salma, Nayanatara
  Prakash, and Mallika Balakrishnan.
\newblock ``{T}hey {L}ook at {V}ulnerability and {U}se {T}hat to {A}buse
  {Y}ou''': {P}articipatory {T}hreat {M}odelling with {M}igrant {D}omestic
  {W}orkers.
\newblock In {\em 31st USENIX Security Symposium}. USENIX Association, 2022.

\bibitem{PSP:Statham2021}
Paul Statham.
\newblock ‘{U}nintended {T}ransnationalism’: The {C}hallenging {L}ives of
  {T}hai {W}omen who {P}artner {W}estern {M}en.
\newblock {\em Population, Space and Place}, 27(5):e2407, 2021.

\bibitem{CHI:StrClaLai19}
Angelika Strohmayer, Jenn Clamen, and Mary Laing.
\newblock Technologies for {S}ocial {J}ustice: {L}essons from {S}ex {W}orkers
  on the {F}ront {L}ines.
\newblock In {\em Proceedings of the 2019 CHI {C}onference on {H}uman {F}actors
  in {C}omputing {S}ystems}, pages 1--14, 2019.

\bibitem{CHI:StrLaiCom17}
Angelika Strohmayer, Mary Laing, and Rob Comber.
\newblock Technologies and social justice outcomes in sex work charities:
  Fighting stigma, saving lives.
\newblock In {\em Proceedings of the 2017 CHI Conference on Human Factors in
  Computing Systems}, pages 3352--3364, 2017.

\bibitem{Thailand:sunanta2014}
Sirijit Sunanta.
\newblock Thailand and the {G}lobal {I}ntimate: {T}ransnational {M}arriages,
  {H}ealth {T}ourism and {R}etirement {M}igration.
\newblock 2014.

\bibitem{QR:trainor2021}
Lisa~R Trainor and Andrea Bundon.
\newblock Developing the {C}raft: {R}eflexive {A}ccounts of {D}oing {R}eflexive
  {T}hematic {A}nalysis.
\newblock {\em Qualitative {R}esearch in {S}port, {E}xercise and {H}ealth},
  13(5):705--726, 2021.

\bibitem{CHI:TFERD21}
Emily Tseng, Diana Freed, Kristen Engel, Thomas Ristenpart, and Nicola Dell.
\newblock A digital safety dilemma: Analysis of computer-mediated computer
  security interventions for intimate partner violence during covid-19.
\newblock In {\em Proceedings of the 2021 CHI Conference on Human Factors in
  Computing Systems}, pages 1--17, 2021.

\bibitem{veena2007revisiting}
N~Veena.
\newblock Revisiting the {P}rostitution {D}ebate in the {T}echnology {A}ge:
  {W}omen who use the{I}internet for {S}ex {W}ork in {B}angkok.
\newblock {\em Gender, {T}echnology and {D}evelopment}, 11(1):97--107, 2007.

\bibitem{warford2022sok}
Noel Warford, Tara Matthews, Kaitlyn Yang, Omer Akgul, Sunny Consolvo,
  Patrick~Gage Kelley, Nathan Malkin, Michelle~L Mazurek, Manya Sleeper, and
  Kurt Thomas.
\newblock Sok: A framework for unifying at-risk user research.
\newblock In {\em 2022 IEEE Symposium on Security and Privacy (SP)}, pages
  2344--2360. IEEE, 2022.

\end{thebibliography}

\appendices

\section{Topic Guide}\label{app:interview-guide}


\subsection*{Topic 1: Background/History}

\begin{itemize}
\setlength\itemsep{0em}
    \item Can you tell me your approximate age and where you are from originally?
    \item How long you have been in Pattaya?
    \item What do you like or dislike about living in Pattaya?
    \item What brought you to Pattaya?
    \item What did you do before you came to Pattaya?
    \item If you work, what do you do now in terms of work?
    \item Have you ``settled'' in Pattaya? Are you planning on staying/making a life/home in Pattaya?
    \item Are you planning to return to your home town? Where do you see yourself over the next few years?
\end{itemize}

\subsection*{Topic 1A: Being a woman in Thailand}

\begin{itemize}
\setlength\itemsep{0em}
    \item What is it like being a woman in Thailand? Why is that your perspective?
    \item What differences have you experienced as a woman moving from a rural community to an urban setting?
    \item What are the primary challenges/barriers for women/opportunities for women in Thailand/Pattaya?
    \item What is your perception of the role of women in Thai society?
    \item What is it that you value most in life? Can you explain.
\end{itemize}

\subsection*{Topic 2: Support Networks/Home}

\begin{itemize}
\setlength\itemsep{0em}
    \item Who do you see as part of your family?
    \item What/where do you consider home?
    \item Who is part of your support network (e.g. friends, family, colleagues, etc.)?
    \item What role do your friends/family members play in your life in terms of support (e.g. emotional support, financial support, etc.)?
    \item What is the role of women’s centres in your life (e.g. what does it give you to come here, how do you feel being/having to come here, why are you here?)
    \item What is the role of the centre’s staff/volunteers (e.g. what support do staff members provide)?
\end{itemize}

\subsection*{Topic 3: Learning at the Centre}

\begin{itemize}
\setlength\itemsep{0em}
    \item Which class/es do you like the most and why?
    \item Which class/es do you find the most useful/helpful and why?
    \item What have you learnt over the last year that you have found most useful and why?
    \item Can you give me an example of where that learning made a difference to you or others in your support network?
    \item How do you think your role as a woman has influenced your learning opportunities (both within the centre and in everyday life)?
\end{itemize}

\subsection*{Topic 4: Information \& Trust}

\begin{itemize}
\setlength\itemsep{0em}
    \item What information do you consider to be personal to you? Can you give me an example.
    \item Thinking about that example, who would you feel comfortable sharing that information with and why? How would you share that information?
    \item How do you decide what information to trust? How do you decide which people to trust?
    \item How do you know what information to share and who to share with?
    \item Can you think of an example where someone has betrayed your trust? Talk through an example -- how did it make you feel, has it changed how you relate to others or to this particular person?
    \item Explain your trust and information sharing with and within the centre (e.g. with staff, among women). What are examples?
\end{itemize}

\subsection*{Topic 5: Digital Technology/Security}

\begin{itemize}
\setlength\itemsep{0em}
    \item How do you communicate with others? (e.g. how do you share information, advice?) Thinking about the example you gave of what you consider personal information, how would you communicate this?
    \item What is your preferred method of communication online/while using digital devices? (e.g. which platforms do you use)? Why do you use these platforms?
    \item What technological devices (e.g. laptops, mobile phones, etc.) do you and members of your household own?
    \item What digital platforms do you use the most, and why? Why do you use these platforms?
    \item What do you like about certain digital technologies?
    \item What do you not like about certain digital technologies?
    \item If you live with others, how is the use of technology shared and/or managed in your household? Please provide examples if possible and describe how this affects your relationship with your family/household.
    \item How do you learn a new technology or digital program? Please provide stories/examples if possible
\end{itemize}

\subsection*{Topic 6: Additional Thoughts}

\begin{itemize}
\setlength\itemsep{0em}
    \item Is there anything else you’ve not shared that you think I should know?
\end{itemize}

\newpage

\section{Participant Tables}\label{app:participant-tables}

\begin{table}[h!]
\centering
  \label{tab:participants-women}
  \resizebox{\columnwidth}{!}{%
  \begin{tabular}{cccc}
    \toprule
    ID & Ages& Home Province/Region & Minutes \\
    \midrule
    P1 + P2  &  53, 48& Udon Thani, Udon Thani & 79  \\
    P3 + P4 & 45, 57& Mae Sai, Chonburi & 50  \\
    P5 + P6 & 28, 30& Isan, Isan & 44\\
    P7 + P8 & 42, 46& Isan, Isan& 63 \\
    P9 -- P11 & 46, 43, 40& Isan, Isan, Sisaket & 69  \\
    P12 & 40 & Isan & 51\\
    P13 + P14 & 51, 44& Isan, Khorat & 65\\
    P15 + P16 & 45, 40& Isan, Isan &  42\\
    P17 + P18& 29, 26& Isan, Isan& 54\\
    P19 + P20& 46, 48& Petchabun, Chonburi& 34\\
    P21& 49& Isan &  33\\
    P22 & 43& Isan & 26\\
    P23 & 37& Bangkok & 53\\
    P24& 50& Udon Thani& 45\\
    P25& 31& Isan & 32\\
    P26 &57& N/A & 22\\
    P27& 34& Chonburi & 27 \\
    P28 &46& Isan& 23\\
    P29 & 46& Isan& 34\\
    P30 & 18& Udan Thani& 69\\
    P31& 52& Chonburi & 38\\
    P32 & 40& Isan & 85\\
    P33 & 44& Udan Thani & 62\\
    P34& 41& Bueng Kan & 105\\
    P35 &57& Udan Thani & 42\\
    P36 & 38 & Isan & 40\\
    P37 & 62& Chonburi & 65\\
    P38 -- P40 & 48, 57, 41& Bueng Kan, Isan, Udan Thani& 57\\
    P41 + P42 &60, 57& Bangkok, Chonburi & 56\\
    P43 & 62& Bangkok & 74\\
    P44 -- P46 &27, 48, 43& Buriram, Phayao, N/A & 56\\
    P47 & 41& Korat & 47\\
    P48& 27& Chonburi & 26\\
    P49& 51& Chonburi & 68\\
    P50 & 54& Khon Kaen & 27\\
    P51&36& Isan& 22\\
    P52& 44& Udan Thani & 31\\
    P53 -- 57 & 48, 38, 52, 48, 57& Udan Thani, Chonburi, Bueng Kan, Korat, Udan Thani& 49\\
    P58 &38& Chonburi & 36\\
    P59 + P60 & 46, 47& Isan, Isan& 51\\
   \bottomrule
 \end{tabular}%
     }  
\caption{\footnotesize Overview of interviewed Thai women (``P'') in Pattaya, their ages, their home province and lengths of the interviews. All interviews were carried out in the centre.}
\end{table}

\begin{table}[h!]
\centering
   \label{tab:participants-staff}
  \begin{tabular}{cccc}
    \toprule
    ID & Age Group & Nationality & Minutes \\
    \midrule
    V1& 20 -- 29 & European &82\\
    V2 & 50 -- 59 & European &62\\
    V3 & 20 -- 29 & European &56\\
    V4 & 70 -- 79& African &60 \\
    S1& 40 -- 49 & Southeast Asian &34\\
    S2 & 50 -- 59 & Southeast Asian &40\\
    S3 & 70 -- 79 & Southeast Asian &50\\
    S4 & 40 -- 49 & Southeast Asian &77\\
    S5 & 40 -- 49 & Southeast Asian &41\\
    E1 & 50 -- 59 & European &91\\
    E2 & 70 --79 & North American &73\\
    E3 & 60 -- 69 & North American &61\\
    E4 & 70 -- 79 & North American &109\\
    E5 & 40 -- 49 & Southeast Asian &21\\
    E6 & 70 -- 79 & European & 49\\
    E7 & 30 -- 39 & North American & 68\\
    \bottomrule
  \end{tabular}
  \caption{\footnotesize Overview of interviewed centre staff (``S''), volunteers (``V'') and external stakeholders (``E'') in Pattaya, including their age group, nationality, and lengths of the interviews. All participants, except for one, identified as female; thus, we do not report gender. Further, because of the small community of staff and volunteers, age group and nationality are used to ensure the confidentiality of all participants.}
  \end{table}

\section{Reflexive Thematic Analysis Tables}\label{app:reflexive-analysis}

\onecolumn

\begin{landscape}
\begin{longtable}[]{@{}
 >{\raggedright\arraybackslash}p{(\columnwidth - 10\tabcolsep) * \real{0.2791}}
 >{\raggedright\arraybackslash}p{(\columnwidth - 10\tabcolsep) * \real{0.2032}}
 >{\raggedright\arraybackslash}p{(\columnwidth - 10\tabcolsep) * \real{0.1751}}
  >{\raggedright\arraybackslash}p{(\columnwidth - 10\tabcolsep) * \real{0.1120}}
  >{\raggedright\arraybackslash}p{(\columnwidth - 10\tabcolsep) * \real{0.1134}}
  >{\raggedright\arraybackslash}p{(\columnwidth - 10\tabcolsep) * \real{0.1172}}@{}}
\toprule
\begin{minipage}[b]{\linewidth}\raggedright
\textbf{Excerpts from Interviews and Fieldnotes}
\end{minipage} & \begin{minipage}[b]{\linewidth}\raggedright
\textbf{Descriptive/ interpretative summary (round 1)}
\end{minipage} & \begin{minipage}[b]{\linewidth}\raggedright
\textbf{Interpretative summary (round 2, group analysis)}
\end{minipage} & \begin{minipage}[b]{\linewidth}\raggedright
\textbf{Codes}
\end{minipage} & \begin{minipage}[b]{\linewidth}\raggedright
\textbf{Initial reflexively constructed category}
\end{minipage} & \begin{minipage}[b]{\linewidth}\raggedright
\textbf{Final reflexively constructed theme, collaborative write-up}
\end{minipage} \\
\midrule
\endhead


``I have a TikTok page where I tell people how to find a date. It’s a free group and I help people find dates for Thai women. When they join me on TikTok I asked them to join a group on Facebook, it is a private group. When these women join, I ask them to join a scamming group. It is like a hotline for scammers...They [the women] wouldn’t know and they would transfer the money [to the scammers]. There are girls who...later find out that they [the men] are scammers. So they take a photo and post it in the group so that everyone knows they [the man] are scammers.'' (Excerpt from P47 transcript); 
The fieldworker witnessed P47 going around the centre before classes and during lunch and sharing her TikTok and Facebook pages. She instructed them how to join and encouraged women to join and explained what the groups were for. Many women appeared to be joining. Women did not seem to struggle with joining Facebook and TikTok groups, but struggled with other digital skills on their mobile phone (e.g. navigation, calendar, email, etc.) (Excerpt from field notes)

& Participants were at financial risk while looking for a \emph{farang} online. Low digital skills posed a security threat to women, so P47 helped encourage information sharing. P47, through her experiences, had learned best practices for avoiding scams. Women believed that sharing experiences of scammers collectively improved their chances of financial security. Information sharing occurred online and in-person. The centre became a place of technological capability building through information sharing of in-person and online resources, rather than the technological classes and training provided by the centre.

& Most participants had material and socioeconomic insecurity. These insecurities initially appeared to be a lack of digital skills, but were a result of material conditions. Women searched for a \emph{farang} to solve these insecurities, despite the security risks involved. Women attempted to mitigate their risks through in-person and online information-sharing. Women actually understood some of the security risks they faced when engaging with \emph{farangs} online and built technology capabilities that specifically served their goals. The centre, also, was helping women to learn the specific capabilities they needed online because staff were grounded in the specific context and challenges of our participants.

& Security practices navigating Pattaya; Authentication of the \emph{farang}; Digital Literacy; Information-Sharing; The Centre's Role in Women's (in)Securities; Relationships between Women at the Centre; Capability Building at the Centre; Trust Relations at the Centre; Appealing to a \emph{farang}; Centre Influencing Relationships with \emph{Farangs}; Technology and Relationships; Digital Trust Networks; Experience-Based Sharing; Scam Detection

& Collective Security; Security and Technology Teaching at the Centre; Scanning Farangs; Protection from Scammers

& Collective Protection from \emph{Farangs} \\

\bottomrule
\end{longtable}
\end{landscape}

\end{document}